\documentclass[opre,nonblindrev]{informs3mod} 
\providecommand{\FUNDING}[1]{}
\RequirePackage{tgtermes}
\RequirePackage{newtxtext}
\RequirePackage{bm}
\RequirePackage{endnotes}

\OneAndAHalfSpacedXII 

\usepackage[ruled, vlined, linesnumbered]{algorithm2e}
\usepackage{algpseudocode}
\usepackage{tikz}
\SetAlCapFnt{\small}
\SetAlCapNameFnt{\small}

\usepackage{natbib}
 \bibpunct[, ]{(}{)}{,}{a}{}{,}%
 \def\bibfont{\small}%

\usepackage[hidelinks]{hyperref}
\usepackage[utf8]{inputenc}
\usepackage[small]{caption}
\usepackage{amsmath}
\usepackage{booktabs}
\usepackage{amsfonts}
\usepackage{color-edits}

\addauthor{mp}{red}

\usepackage[capitalize,noabbrev]{cleveref}
\crefname{appendix}{Appendix}{Appendices}

\newcommand{\dg}[0]{\mathsf{DG}}
\newcommand{\zero}[0]{\mathbf 0}
\newcommand{\supp}[0]{\mathrm{supp}}
\newcommand{\opt}[0]{\mathsf{OPT}}

\newcommand{\ev}[2]{\mathbb E_{#1} \left [ #2 \right ]}
\newcommand{\one}[0]{\mathbf 1}
\newcommand{\tr}{\operatorname{tr}}
\newcommand{\diag}{\operatorname{diag}}

\newcommand{\sign}{\operatorname{sign}}
\newcommand{\xpar}[1]{\noindent \textbf{#1.}}
\newcommand{\cU}[0]{\mathcal U}
\usepackage{pifont}
\newcommand{\cmark}{\ding{51}}
\newcommand{\xmark}{\ding{55}}
\usepackage{newfloat}
\usepackage{float}
\usepackage{listings}
\usepackage{booktabs}
\usepackage{subfigure}

\usepackage{caption}
\EquationsNumberedThrough    

\TheoremsNumberedThrough     
\ECRepeatTheorems  %

\MANUSCRIPTNO{} 

\begin{document}


\RUNAUTHOR{Papachristou and Kleinberg}

\RUNTITLE{Measuring Disparity in Social Consensus}

\TITLE{Quantifying and Mitigating Consensus Disparity in Social and Information Networks}

\ARTICLEAUTHORS{%
\AUTHOR{Marios Papachristou}
\AFF{Department of Information Systems, W.P. Carey School of Business, Arizona State University \EMAIL{mpapachr@asu.edu}}

\AUTHOR{Jon Kleinberg}
\AFF{Department of Computer Science,
Cornell University, \EMAIL{kleinberg@cornell.edu}}
} 

\ABSTRACT{%
We introduce a computational framework to measure and optimize disparity, which corresponds to the difference in consensus outcomes attributable to distinct social groups, under classical models of opinion dynamics. We study this problem in the Friedkin-Johnsen setting under uncertainty about group structure and characterize its algorithmic complexity. For the structural analysis, we demonstrate that disparity can be arbitrarily larger than polarization in well-connected networks that nonetheless carry an identifiable group structure. For the mitigation problem, we derive robust formulations and active set optimization procedures to minimize worst-case disparity via recommendation reweighing and opinion seeding. Our methods provide provable guarantees and are validated on multiple real-world social networks. The results bridge opinion dynamics and network optimization, offering computational tools for analyzing and reducing polarization in social networks.


}

\FUNDING{Supported in part by a Simons Investigator Award, a Vannevar Bush Faculty Fellowship, AFOSR grant FA9550-19-1-0183, a Simons Collaboration grant, and a grant from the MacArthur Foundation. The authors would like to thank Panagiotis Adamopoulos, Olivia Liu Sheng, and M. Amin Rahimian for their feedback and suggestions on the paper.}




\maketitle



\section{Introduction}

Online platforms, organizations, and institutions increasingly rely on networked interactions to aggregate public opinion and achieve outcomes through social contagion. In many such settings, participants are heterogeneous and unevenly represented: groups vary in size, activity, structural position, and susceptibility to influence, so a system may appear to reach consensus while masking a more subtle fragility, namely, that the outcome may depend critically on the groups themselves. A central concern in such systems is \textit{polarization}, that is, how far apart opinions are across users or groups, a measure that has proven valuable for understanding social fragmentation and the effects of network structure, as underlined by theories such as the \textit{filter-bubble theory} \citep{pariser2011filter,hosanagar2014will}.

However, polarization alone does not capture \textit{which} groups drive the resulting consensus. In particular, a system may exhibit relatively low opinion dispersion while remaining highly sensitive to the participation of specific groups, due to structural asymmetries in the network.

To illustrate, consider the well-studied political blogs, henceforth \textit{polblogs}, network from \citet{adamic2005political}, which has a well-defined, almost-balanced liberal-conservative structure. Under a natural choice of private opinions, standard polarization measures developed in the literature \citep{musco2018minimizing,racz2023towards,gaitonde2020adversarial} indicate surprisingly relatively low polarization across the network. However, when we examine the dependence of the consensus outcome on each group, an important contrast emerges: the difference is large, indicating that the consensus is highly sensitive to the group structure itself. In other words, while opinions appear moderately aligned, the aggregate outcome is effectively dominated by one group over the other.

A system with low polarization might be perceived as more stable, yet high disparity exposes inherent systemic fragility: changes in participation or coordinated activity from a specific group can substantially shift the consensus. This motivates a complementary measure that captures group-dependent influence on consensus outcomes, not merely the dispersion of opinions, which can be utilized to assess representational fairness, robustness, and resistance to adversarial manipulation.

This concern has become especially salient in real-world settings. Investigations into online political discourse have shown that relatively small, coordinated groups can exert outsized influence on perceived public opinion, as observed during the 2016 U.S. elections and subsequent coordinated campaigns on major social platforms \citep{allcott_social_2017}. Similarly, platform disclosures regarding information operations during events such as the Hong Kong protests in 2019 highlight how consensus narratives can shift dramatically depending on which communities are amplified or suppressed \citep{twitter2019hongkong}. These cases show that consensus outcomes are shaped as much by which groups carry the most weight as by how polarized opinions are. This raises a fundamental managerial and policy question: 

\medskip

\begin{quote}
\emph{\textbf{(RQ1)} How robust is consensus to shifts in group-level participation and influence?} 
\end{quote}

\medskip

A consensus outcome that reflects broad integration across groups is qualitatively different from one that would change substantially if a particular constituency were amplified, disengaged, or strategically manipulated. Yet existing measures of polarization and disagreement \citep{musco2018minimizing,axelrod_preventing_2021} provide limited visibility into this distinction.

\begin{figure}[t]
    \centering
    \includegraphics[width=0.8\linewidth]{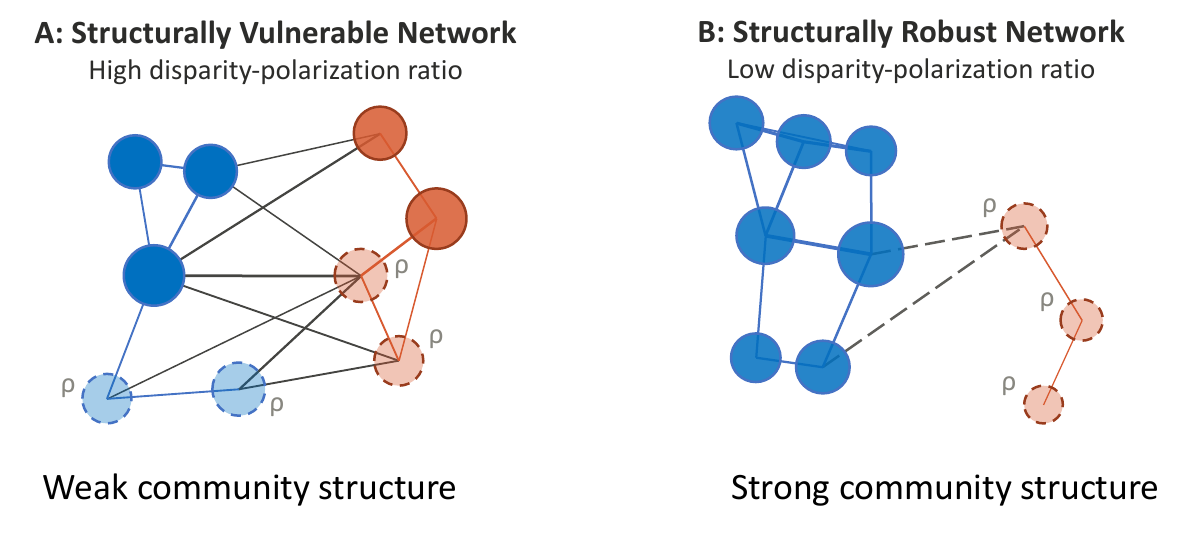}
    \caption{Consensus disparity versus polarization. Node color indicates group membership (the ``blue'' and ``red'' groups); dashed borders mark users whose membership the platform infers with classifier error $\rho$; solid edges are within-group ties, and dashed gray edges are weak cross-group ties. \textbf{(A)} A well-connected network: opinions converge to a concentrated consensus with low polarization that can nonetheless be disproportionately driven by one group, so disparity is large relative to polarization. \textbf{(B)} A network with separated communities: the variance of the opinions is high, so polarization is high and reflects the group structure more directly. Disparity thus exposes the group-dependence of consensus, which cannot be audited by the polarization metric, as in~(A).}
    \label{fig:illustration}
\end{figure}

Motivated by this phenomenon, we introduce \textit{consensus disparity}, a \textit{counterfactual} measure that quantifies how differently a network would converge if private opinions from distinct groups were considered in isolation: given a partition into two groups, we compare the consensus outcome attributable to one group's private opinions with that attributable to the other's. A large disparity indicates that consensus is highly sensitive to group identity, whereas a small disparity suggests opinions are well-integrated and outcomes are robust to shifts in participation. Managerially, disparity captures a form of \textit{systemic risk}: high disparity means small shifts in participation or targeted amplification can significantly alter the outcome, while low disparity indicates a broadly representative and stable consensus. \cref{fig:illustration} illustrates the metric and how it complements polarization.

In practice, though, platforms do not control users' private opinions or group memberships, but they do control interaction structure through recommendation policies and content exposure. The platform must therefore intervene in the face of uncertainty about the very quantity it is trying to manage. This leads to our central artifact design question:

\begin{quote}
  \emph{\textbf{(RQ2)} How should a platform modify its interaction network to minimize the sensitivity of consensus outcomes to group structure, when group membership is only partially or noisily observed?}
\end{quote}

We situate our study of the disparity under the well-established social learning model of \citet{Friedkin1990}, modeling platform interventions as controlled changes to network links (recommendation reweighing) and to a subset of users' expressed opinions (opinion seeding). In both cases, the platform operates against an adversarial group structure drawn from an uncertainty set around its nominal estimate. This robust formulation captures the realistic operating condition of a platform: not the case where group labels are known exactly, but the case where they are estimated and potentially manipulated.

We make two main contributions: First, we introduce the disparity metric, a \textit{complementary metric to polarization} that quantifies how differently a network would converge if distinct groups drove influence. Unlike polarization, disparity captures the dependence of consensus outcomes on group structure, revealing asymmetries in influence invisible to existing measures. Then, we develop a general formulation based on the platform's uncertain opinions about the group structure. When the platform's opinion aligns with the structural separation into two groups, we show that disparity can be significantly higher than the polarization, signaling a high amount of systemic risk. Second, we study \textit{how} platforms can mitigate disparity under uncertainty about group structure by modeling their opinions, accounting for uncertainty, and by designing targeted interventions, including robust recommendation reweighing and opinion seeding.
 
These results establish a unified framework for measuring and mitigating group-dependent influence in social networks under uncertainty, with direct implications for platform design and governance. Specifically, we show that even though the platform may face high uncertainty about the group structure, our proposed methods minimize disparity while incurring \textit{minimal change in the platform's microstructure}, making the approach realistic and feasible in real-world settings.


The remainder of the paper is organized as follows. In \cref{sec:related_work}, we give an overview on the literature on opinion dynamics, polarization, and platform interventions. In \cref{sec:disparity}, we introduce the opinion dynamics models and notation used throughout the paper, define the consensus disparity metric, and present its general formulation based on a group structure, along with an illustrative example on a real-world network that highlights its distinction from polarization. In \cref{sec:interventions}, we study platform interventions to minimize the worst-case value of the disparity. \cref{sec:experiments} presents experimental results on real-world networks. Finally, \cref{sec:discussion} discusses potential extensions of the framework and concludes with managerial implications and future directions.


\section{Related Work} \label{sec:related_work}

Our work contributes to the growing literature on polarization, disagreement, and strategic manipulation in online platforms \citep{Chen2022,gaitonde2020adversarial,musco2018minimizing,zhu2021minimizing,Chitra2020,wang2024relationship,racz2023towards}, as well as to the broader literature on platform governance and interventions. The most closely related studies are \citet{Chen2022} and \citet{musco2018minimizing}, which introduce measures of polarization and disagreement within the Friedkin-Johnsen (FJ) model \citep{Friedkin1990} and analyze how the platform can mitigate these quantities; follow-up works extend this line by considering opinion seeding and link manipulation \citep{wang2024relationship,racz2023towards}. Our work builds on this literature by introducing a group-aware disparity metric and studying how it can be optimized via platform interventions under the FJ model. Because the metric is formulated in terms of a group structure, the resulting link manipulation and seeding problems are richer than their polarization counterparts in \citet{musco2018minimizing}, and naturally admit robust formulations connected to robust eigenvalue optimization \citep{boyd2004convex} and robust submodular optimization \citep{krause2008robust,kaya2025randomized}. Our approach is also related to network optimization problems in adjacent domains, including selection of informative review subsets \citep{zhang2021review} and influence maximization and control \citep{Kempe2015,gunnecc2020least,raghavan2022influence,raghavan2022rapid,han2023cost}.

Beyond the algorithmic literature, our work connects to platform economics on how digital platforms shape information flows and visibility \citep{oestreicher2012visible,susarla2012social,liu2020go}, and how algorithmic curation can exacerbate polarization \citep{hosanagar2014will}. Our paper is also related to the social science literature on political polarization and echo chambers \citep{garimella2017reducing,axelrod_preventing_2021} and to work documenting how recommendation algorithms amplify ideological segregation \citep{pariser2011filter,hosanagar2014will,adamic2005political}. By providing a formal way to quantify how consensus outcomes differ across groups, our framework bridges algorithmic models of social influence and empirical studies of polarization, offering a new lens for analyzing how structural and intrinsic factors jointly shape disparities in consensus formation.

\section{The Disparity Metric} \label{sec:disparity}

Online platforms shape opinion formation by controlling patterns of interaction. Even when individuals' private opinions are fixed, network structure can cause different groups to converge on different consensuses. To capture this phenomenon, we introduce the disparity metric, which quantifies how network-mediated opinion dynamics amplify differences between groups. \cref{tab:summary} shows a summary of the metrics and interventions studied in this paper. 

\begin{table}[t]
    \centering
    \scriptsize
    \begin{tabular}{p{4cm}p{2cm}p{2cm}p{3cm}p{2cm}}
    \toprule
    Metric & Group structure ($C$) is unknown & Intrinsic opinions ($s$) are unknown & Recommendation reweighing & Opinion seeding \\
    \midrule
    Polarization $\mathcal P(s, L)$ & n/a & n/a & \citet{musco2018minimizing,racz2023towards} & \citet{musco2018minimizing} \\
    Conditional disparity $f(s, L, A)$ & \xmark & \xmark & \cref{sec:link_recommendation} & \cref{sec:seeding}, and \cref{sec:opinion_seeding} \\
    Average disparity $g(s, L, C)$ & \cmark & \xmark & \cref{sec:link_recommendation} & \cref{sec:seeding}, and \cref{sec:opinion_seeding} \\
    Structural disparity $h(L, C)$ & \cmark & \cmark & \cref{sec:link_recommendation} & \cref{sec:seeding}, and \cref{sec:opinion_seeding} \\
    \bottomrule
    \end{tabular}
    \caption{Metrics and interventions studied in this paper.}
    \label{tab:summary}
\end{table}

According to the FJ model, we let $G=(V,E)$ be a weighted, undirected graph with $n=|V|$ users and weights $w : V \to \mathbb R_{\ge 0}$ with $w_{ii} = 1$. Let $d_i$ be the weighted degree of user $i$ and let $m$ be the sum of the edge weights, i.e., $m = \frac 1 2 \sum_{i = 1}^n d_i$.  Each user $i$ has a private opinion $s_i \in \mathbb{R}$. We assume that the private opinion vector is normalized, that is, $\| s \| = 1$ and is zero mean, i.e., $\one^T s = 0$, with $\one$ being the all-ones column vector. In the FJ model, each agent incurs a quadratic cost for failing to reach consensus with her neighbors and her private opinion \citep{Bindel2015}, yielding each user $i$ iteratively exchanging opinions with their neighbors: 

\begin{equation*}
    x_i(t + 1) = \frac {\sum_{j \neq i} w_{ij} x_j(t) + s_i} {1 + \sum_{j \neq i} w_{ij}},
\end{equation*}

which yields the following equilibrium as $z = \lim_{t \to \infty} x(t)$:  $$z = (I + L)^{-1} s,$$ where $L = D - W$ is the Laplacian matrix of $G$. Traditionally, platforms have been interested in measuring polarization. To make this concrete, we let $\bar z = \frac 1 n \sum_{i = 1}^n z_i$ be the average expressed opinion at equilibrium. The polarization measures the deviation from $\bar z$, that is $$\mathcal P(s, L) = \|  z- \bar z \one \|^2 = s^T (I + L)^{-2} s.$$

However, platforms may also want to measure whose voice gets amplified. To achieve that, we will gradually proceed with building the definition of the disparity metric we introduce in this paper. We begin with a setting in which the population is partitioned into two non-overlapping groups\footnote{Our analysis can extend to multiple groups and overlapping groups; see our discussion in \cref{sec:discussion}.}, $A$ and $\bar A = V \setminus A$. Such partitions naturally arise when platforms have access to group information, e.g., demographic categories, political affiliations, experimental cohorts, or known communities. In these cases, it is meaningful to ask how differently opinions would converge if influence were driven primarily by one group versus the other.

Given the partition $(A, \bar A)$ of the users we define the \textit{conditional disparity} as the difference between the consensus outcomes $z_A$ and  $z_{\bar A}$ attributable to groups $A$ and $\bar A$ under the FJ model, namely, $$f(s, L, A) = \| z_A - z_{\bar A} \|^2  \text{ where, }   z_A = (I + L)^{-1} \underbrace{(\one_A \odot s)}_{s_A},   \text{ and }   z_{\bar A} = (I + L)^{-1} \underbrace{(\one_{\bar A} \odot s)}_{s_{\bar A}}.$$

An alternative way to view the conditional disparity is through the introduction of an assignment vector $r \in \{ +1, -1 \}^n$ with entries $r_i = +1$ if $i \in A$ and $r_i = -1$ if $i \in \bar A$. Letting $y = s \odot r$ be the vector of opinions after applying the assignment $r$, we can show that $$f(s, L, A) = y^T (I + L)^{-2} y.$$

One challenge with the above definition of the conditional disparity is that the platform, in general, lacks knowledge of the group structure and cannot calculate $r$ or, subsequently, $y$ exactly. In the sequel, we relax this assumption and obtain a network-level measure that is independent of a particular partition, which captures the platform's uncertainty about the group structure, assuming that the assignments to groups are random. To remove the dependence on a specific assignment $r$, we average the conditional disparity over randomized group assignments. Specifically, we assume that the entries $r_i$ of the assignment vector $r$ are Rademacher variables that are correlated with $C_{ij} = \ev {} {r_i r_j}$ for $i \neq j$ and $C_{ii} = \ev {} {r_i^2} = 1$. We define the \textit{average disparity}, or, henceforth \textit{``disparity''}, to be the disparity under the correlation matrix $C$ that governs this group structure, namely $$g(s, L, C) = \ev {A \sim C} {f(s, L, A)}, \quad \text{with} \quad C \succeq 0,   \diag(C) = \one.$$

\noindent By expanding the expectation, we get that

\begin{equation*} 
    g(s, L, C)= \ev {A \sim C} {f(s, L, A)} = \sum_{i = 1, j = 1}^n m_{ij} C_{ij} s_i s_j = s^T Z s \quad \text{for} \quad Z = M \odot C. 
\end{equation*}

Note that if we let $C$ be the matrix of all-ones, that is $C = \one \one^T$, then the disparity metric reduces to the polarization metric, i.e., $g(s, L, \one \one^T) = \mathcal P(s, L)$. In general, note that relying on the correlation structure matrix $C$ has several benefits, as it can encode several settings that are managerially relevant and realistic to modern platforms. In the next section, we present an example illustrating the managerial relevance of the disparity metric.

Finally, the disparity metric $g(s, L, C) = s^T Z s$ depends on the private opinion vector $s$, and therefore measures a joint property of the network structure and the current opinion distribution. This dependence is natural: a platform that intervenes in real time observes its users' current opinions and should optimize for the specific disparity induced by those opinions. However, in settings where the platform wishes to design interventions that are robust to any opinion distribution, for instance, when opinions are unobservable, adversarially chosen, or likely to shift over time, it is more appropriate to consider the \emph{structural disparity}, defined as the worst-case disparity over all normalized opinion vectors: $$h(L, C) = \max_{s :\|s\| = 1} g(s, L, C) = \lambda_{\max}(Z).$$

The structural disparity depends only on the network $L$ and the group structure $C$, and captures the maximum sensitivity of consensus outcomes to group structure across all possible opinion distributions. Our intervention design problems are targeted towards the structural disparity. 

\xpar{Axiomatization of the disparity metric} Our definitions of $f(s, L, A)$, $g(s, L, C)$, and $h(L, C)$ rest on structural assumptions about how to attribute influence to each group. When asking \textit{``how much does group $A$ contribute to the consensus,''} a natural starting point is the Shapley attribution literature \citep{shapley1953value}: we credit each group with the portion of the consensus outcome attributable to its private opinions, holding the network fixed. This raises a decomposition question: given $z = (I + L)^{-1} s$, how do we split $z$ into parts attributable to $A$ and $\bar A$?
The most natural decomposition follows from the linearity of the FJ model. Since $s = s_A + s_{\bar A}$ where $s_A = \one_A \odot s$ and $s_{\bar A} = \one_{\bar A} \odot s$, linearity of the map $(I + L)^{-1}$ gives, $z = z_A + z_{\bar A}$. 

This decomposition is additive and exhaustive: the two attributions sum exactly to the true consensus, with no residual, and it requires no assumption about what the other group ``would have said'' in a counterfactual world. Users of both groups remain present in the network and continue to propagate influence; the decomposition separates only the sources of the private-opinion signal, not the agents themselves.

One might ask whether this decomposition is the right one, for instance, whether at the measurement of $z_A$ one should inject an opinion $\alpha s_i + \beta$ to the other group, or remove its users entirely. In \cref{app:definition} we show that this decomposition is the unique one satisfying three properties any reasonable attribution should have: \textit{(i)} additivity, the two attributions sum to the true consensus; \textit{(ii)} affine invariance, the measured disparity is unchanged if we re-scale or re-center the other group's opinions, which a platform cannot observe; and \textit{(iii)} linearity in the opinions. Consequently, the disparity is exactly invariant to replacing the other group's opinions by any constant $\beta$, so we choose $\alpha = 1, \beta = 0$ without loss of generality. Removing the other group's users entirely would instead change the Laplacian $L$ and violate additivity, making the metric depend on an arbitrary subgraph rather than a structural property of the full network.

\subsection{Managerial Example: Disparity and Polarization on the Polblogs Network} \label{sec:managerial_example}

\begin{figure}[t]
    \centering
    \includegraphics[height=1.95in]{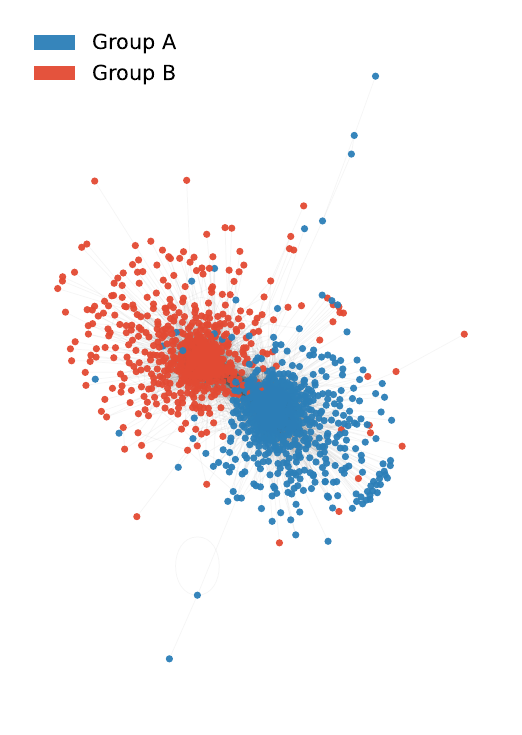} \quad
    \includegraphics[height=1.95in]{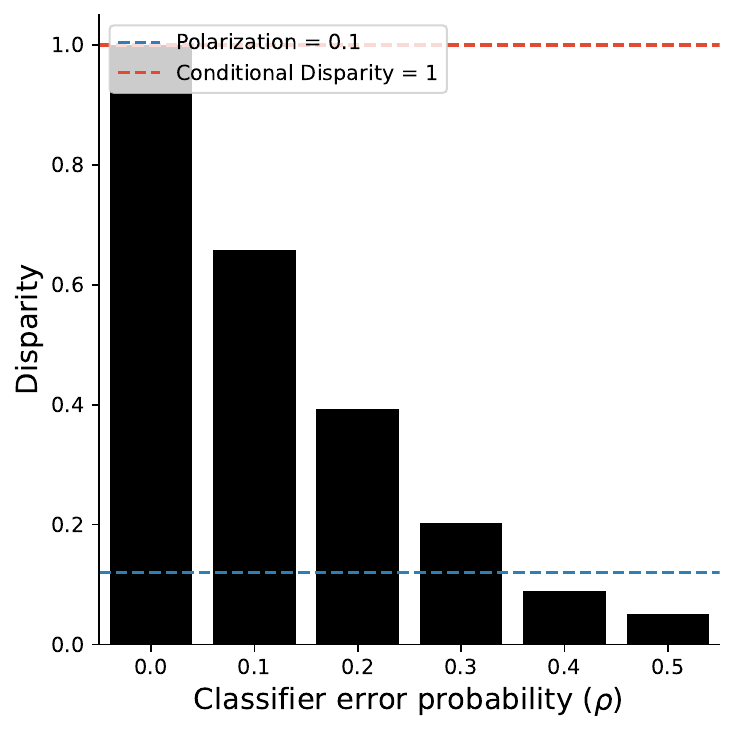}
    \caption{Left: The Polblogs network with the nominal group structure $\bar r$. Right: Values of the disparity $g(s, L, C(\rho))$ for various values of the platform's classifier error probability $\rho$ for the polblogs dataset (Left), which induces a group structure $C(\rho)$. 
    }
    \label{fig:example}
\end{figure}

In 2004, the political blogosphere was one of the primary arenas of online political discourse, with thousands of blogs linking to one another to amplify,  rebut, or contextualize political arguments~\citep{adamic2005political}. Platform designers and researchers of that era were primarily concerned with whether the network was \emph{polarized}, that is, whether liberal and conservative blogs were talking past each other. However, polarization alone does not reveal a complementary and arguably more urgent concern: even in a network that appears moderately integrated by polarization measures, the aggregate consensus can be entirely dominated by one ideological cluster. As we show below, the disparity metric surfaces exactly this blind spot.

To illustrate the complementarity between disparity and polarization, we consider the aforementioned political blogs (Polblogs) network, a widely studied benchmark with a well-defined partition into liberal and conservative communities \citep{adamic2005political}, where the liberal (resp. conservative) blogs are assigned a private opinion $s_i = +1 / \sqrt n$ (resp. $s_i = -1 / \sqrt n$). 

We first evaluate a standard polarization measure $\mathcal P(s, L) = g \left ( s, L, \one \one^T \right )$, which captures the dispersion of opinions in the network, and assumes that the platform is aware of the ideological position ($s_i$) of each user. Surprisingly, in this instance, even if it visually seems that the network is highly polarized, polarization is relatively low and equals 0.1, indicating that the equilibrium opinions do not vary much.

However, in practice, the platform may not be able to measure polarization directly and instead relies on noisy estimates $r$ of the labels, which in general may differ from the nominal group structure $\bar r$. 

To put this in context, we assume that the platform has access to a predictive model $\hat {f} $ that, given information about each user $i$, produces an estimate $r_i = \hat f(x_i)$ of their membership based on their features $x_i$. For simplicity, we assume that the platform treats users independently (the case where pairs of users are considered is also straightforward under our framework), and that it possesses a classifier which has accuracy $1 - \rho$, where $\rho \in [0, 1/2]$. This means that a user with nominal membership $\bar r_i$ is misclassified ($r_i = - \bar r_i$) with probability $\rho$ and is correctly classified with probability $1 - \rho$. 

First, assuming an ideal classifier ($\rho = 0$), we evaluate the disparity metric, and a very different picture emerges: the disparity is significantly higher than the polarization, indicating that the resulting consensus is highly sensitive to which group's opinions are present. Specifically, under the nominal group structure $\bar C = \bar r \bar r^T$, the disparity achieves its universal upper bound and equals 1, i.e., $g(s, L, \bar C) = 1$, which is 10 times higher than polarization, meaning that the voices of one ideological group dominate the consensus on expectation.


If the classifier has a non-zero error $\rho$ uniformly across the users, we can calculate the group structure of the random partition $r$ to have entries $C_{ij}(\rho)$ given by:

\begin{align} \label{eq:correlation_matrix}
    C_{ij}(\rho) = \ev {} {r_i r_j} = 
    \begin{cases}
        1, & i = j \\
        (1 - 2 \rho)^2 \bar r_i \bar r_j, & i \neq j 
    \end{cases}.
\end{align}

In \cref{fig:example}, we report the value of the disparity for various values of $\rho$ ranging from $\rho = 0$ to $\rho = 0.5$, and observe that the disparity can take vastly different values as a function of $\rho$; for instance, for $\rho = 0.1$ the disparity is around $6 \times$ higher. This example exposes both a limitation of the polarization metric and the flexibility of the more general disparity measure. 

While polarization suggests that the system exhibits surprisingly low opinion variance, disparity reveals an asymmetry in influence across groups. Thus, a system with low polarization may still be highly vulnerable to its group structure, and disparity provides a diagnostic that captures this group-dependent sensitivity, which traditional polarization measures do not reflect.
This can be further shown in the general case: specifically, we can show that in a well-connected network with an identifiable group structure, the gap between the disparity and polarization can grow large, and in fact grows with the second smallest eigenvalue ($\lambda_2$) of the Laplacian, i.e., its algebraic connectivity:

\begin{proposition}
\label{theorem:gap}
Let $G = (V, E)$ be a connected weighted network with Laplacian $L$ and eigenvalues $0 = \lambda_1 < \lambda_2 \le \cdots \le \lambda_n$. Suppose private opinions are aligned with the nominal group membership, that is $s_i = \bar{r}_i/\sqrt{n}$ where $\bar{r}_i = +1$ for $i \in A$ and $\bar{r}_i = -1$ for $i \in \bar{A}$, and assume the partition is balanced so that $\sum_i \bar{r}_i = 0$. Assume that the platform possesses a predictive model with error $\rho$, that is $\Pr [r_i \neq \bar r_i] = \rho$ with $\rho \in [0, 1/2)$, and corresponding group structure $C(\rho)$ (\cref{eq:correlation_matrix}). If $R(\rho, \lambda_2) = \frac{g(s, L, C(\rho))}{g(s, L, \one \one^T)}$ is the ratio between the disparity and the polarization, then: $$R(\rho, \lambda_2) \ge {(1 - 2\rho)^2} {(1 + \lambda_2)^2}.$$
\end{proposition}

For instance, in complete networks where $\lambda_2 = n$, the gap can grow unbounded regardless of $\rho$. As both the example and \cref{theorem:gap} show, the degree of uncertainty about the partitions can significantly affect outcomes; for that reason, platforms may want to design interventions that are robust to uncertainty ($\rho$). The platform can intervene in two ways to mitigate disparity: First, it may modify the recommendation link weights to minimize disparity subject to some bounded uncertainty, and, secondly, it may be able to select a set of users to intervene with the goal of minimizing the worst-case disparity as well.

\section{Platform Interventions under Uncertainty} \label{sec:interventions}

Online platforms rarely have direct control over users' private opinions or group memberships. Their primary intervention mechanisms are recommendation reweighing and opinion seeding. Since the correlation structure $C$ is estimated from noisy behavioral signals and private opinions $s$ may be unobservable or adversarially perturbed, we formulate both intervention problems directly in their robust form. 

For simplicity, we will assume the setting of \cref{sec:managerial_example} where the platform possesses a predictive model $\hat f$ that predicts the nominal membership of a user $i$ with error $\rho$, for some $\rho \in [0, 1/2]$ that is $\Pr_{r_i \sim \hat f} [r_i \neq \bar r_i] = \rho$. According to this simple model, the non-diagonal entries of the resulting group structure $C$ are $C_{ij} = (1 - 2 \rho)^2 \bar r_i \bar r_j$ and satisfy $| C_{ij} - \bar C_{ij} | = \left | (1 - 2 \rho)^2 \bar r_i \bar r_j - \bar r_i \bar r_j  \right | = |4 \rho^2 - 4 \rho| | \bar r_i \bar r_j | = 4 \rho (1 - \rho)$. 
Thus, it is reasonable for the platform to construct an uncertainty set $\mathcal U(\rho)$ (which, for brevity, we will denote by $\mathcal U$ whenever it is evident from the context) around the nominal estimate $\bar C$: $$\mathcal U(\rho) = \left \{ C \in \mathbb R^{n \times n} : C \succeq 0, \diag (C) = \one, | C_{ij} - \bar C_{ij} | \le 4 \rho (1 - \rho),   \forall i \neq j \right \}.$$

We note that $\mathcal U(\rho)$ is a conservative relaxation of the parametric family $\{C(\rho') : \rho' \le \rho\}$ generated by the uniform independent flip model of \cref{sec:managerial_example}: under that model every off-diagonal entry satisfies $|C_{ij} - \bar{C}_{ij}| = 4\rho(1-\rho)$ exactly, whereas $\mathcal U(\rho)$ is convex, tractable, and richer as a feasible set for the adversarial problem, and the robust guarantee holds a fortiori for the flip model. In \cref{sec:experiments} we use this relaxation, since we train classifiers on graph-based embeddings that are not independent across users.

These problems differ structurally from the closest prior work: \citet{musco2018minimizing} and \citet{racz2023towards} minimize polarization and disagreement, quantities with no group structure. The group structure changes the character of both problems, since per-edge contribution to the disparity now depends on $C$, so which edges to reweigh is sensitive to the platform's noisy beliefs about group structure, which informs our algorithm design.

\subsection{The Recommendation Reweighing Problem} \label{sec:link_recommendation}

We model platform interventions as controlled modifications to the network's edge weights, representing changes in exposure, recommendation strength, or interaction intensity. Since the platform operates under uncertainty over both the correlation structure $C$ and the private opinions $s$, the natural objective is to minimize worst-case disparity over all plausible group structures and opinion vectors simultaneously. Formally, denoting by $L$ the convex feasible set of trace-constrained Laplacians $\mathcal L = \{ L^{n \times n} \text{ Laplacian} : \tr(L) = 2m, 0 \le w_e \le \bar w   \forall e \in E \}$. The platform's primary problem is to solve

\begin{align} \tag{R-Link} \label{eq:r_link}
    \min_{L \in \mathcal L} \max_{C \in \mathcal U} \max_{s : \| s \| = 1} g(s, L, C).
\end{align}

This formulation captures the realistic operating condition of a platform: it selects a link structure that performs well not for a single estimated value of $C$ and $s$, but against the worst-case group structure and opinion distribution within the uncertainty set $\mathcal U$.

On the other hand, the case where $\rho = 1/2$ is also interesting, as the constraint $|C_{ij} - \bar C_{ij}|$ becomes trivial (upper bounded by 1). In that case, the optimization is over the set of unconstrained correlation matrices $\{ C \in \mathbb R^{n \times n} : C \succeq 0, \diag(C) = \one \}$. In that case, the extremal value of the disparity is obtained when $C = \one \one^T$, and the partition is assumed to be balanced, the problem reduces to maximizing the algebraic connectivity of the network.

In the general case, directly optimizing for $\rho \in (0, 1/2)$ \eqref{eq:r_link} has two problems: First, $g(s, L, C)$ is non-convex in $L$ due to the matrix function $M(L) = (I + L)^{-2}$, which makes the outer minimization intractable in general. Second, even if convexity were available, the inner maximization over $C$ does not have an analytical solution for general $\rho$. We address both obstacles in turn: First, to handle non-convexity in $L$, we replace $g$ with the surrogate disparity $\tilde g(s, L, C) = s^T(X \odot C)s$, for $X = (I + L)^{-1}$, which extends the sum of polarization and disagreement metric of \citep{musco2018minimizing} by incorporating the influence of the group structure. The surrogate function $\tilde g$ serves as an upper bound to the disparity metric $g$ for any set of initial opinions $s$, Laplacian $L$, and group structure $C$, so minimizing the surrogate $\tilde g$ provides a meaningful upper bound for $g$. Additionally, we note that $\tilde g$ is matrix-convex in the Laplacian $L$ (see \citet{musco2018minimizing}). 

\begin{proposition}
\label{prop:surrogate_upper_bound}
For every undirected weighted connected network $G$,  group structure $C \succeq 0$, and set of initial opinions $s$, the surrogate disparity $\tilde g(s, L, C) = s^T (X \odot C) s$ is an upper bound to the true disparity $g(s, L, C) = s^T (M \odot C) s$ and satisfies $0 \le \tilde g(s, L, C) - g(s, L, C) \le \mu_{\max}$ where $\mu_{\max} = \max_{i \in [n]} \frac {\lambda_i(L)} {(1 + \lambda_i(L)))^2} \le \frac 1 4$. The surrogate disparity $\tilde g(s, L, C)$ is matrix-convex in the set $\mathcal L$.
\end{proposition}

Thus, the platform can consider the alternative minimization problem on the surrogate disparity, which is a convex problem in $L$, and this admits a global minimum $\min_{L \in \mathcal L} \max_{C \in \mathcal U} \max_{s : \| s \| = 1} \tilde g(s, L, C)$. Substituting $\tilde g$ for $g$ and applying the eigenvalue characterization of the worst case over $s$ yields the tractable robust surrogate problem:
  
\begin{align} \tag{RS-Link} \label{eq:rs_link}
    \min_{L \in \mathcal L} \;   \phi(L), \quad \text{where} \quad \phi(L) = \max_{C \in \mathcal U} \; \lambda_{\max}(X \odot C).
\end{align}

One approach to \eqref{eq:rs_link} is to develop an SDP relaxation of \eqref{eq:rs_link}; however, the relaxation cannot scale to large real-world networks. To address the scalability problems in real-world networks, we develop a scalable active set algorithm for large networks. 

\xpar{Leverage scores and gradient updates}  The first component of the active set algorithm includes selecting which edges to update. To achieve that, we first characterize the sensitivity of $\tilde g$ to individual edge weight changes. For a fixed $C$, the gradient of $\tilde g$ with respect to the weight of edge $e = (i, j)$ defines the surrogate leverage of that edge: 


\begin{proposition} \label{theorem:fj_model_link_update}
    In the FJ model, let $\phi(L) = \max_{C \in \mathcal{U}} \lambda_{\max}(X \odot C)$ with $X = (I+L)^{-1}$ be the robust surrogate objective of \emph{(RS-Link)}. For each edge $e \in E$, the gradient of $\phi$ with respect to the edge weight $w_e$ is:
\begin{equation*}
    \frac{\partial \phi}{\partial w_e} = -\tilde{\beta}^{\emph{rob}}_e, \quad \text{where} \quad \tilde{\beta}^{\emph{rob}}_e = v^T (\tilde{Q}_e \odot C^*)v = x_e^T (C^* \odot vv^T) x_e,
\end{equation*}
where $\tilde{Q}_e = Xb_e b_e^T X$, $b_e$ is the incidence vector of edge $e$, $x_e = Xb_e$, $C^* = \operatorname*{arg max}_{C \in \mathcal{U}} \lambda_{\max}(X \odot C)$ is the worst-case group structure, and $v$ is the leading eigenvector of $X \odot C^*$. 
\end{proposition}

Two features of \cref{theorem:fj_model_link_update} are worth highlighting in contrast to the polarization metric: First, both leverage scores depend on $C^*$ through the Hadamard product with $Q_e$ (resp. $\tilde Q_e$), meaning that the relative value of reweighting an edge is sensitive to the platform's opinions about group structure. An edge that is highly beneficial under one group structure may be neutral or even harmful under another. This is precisely why the nominal version of the problem, optimized for a single nominal group structure $\bar C$, can produce poor interventions when group structure is uncertain.

\xpar{An active set algorithm for large networks} We develop a scalable active set algorithm that alternates between two steps: minimizing the surrogate disparity over $L$ for a fixed worst-case group structure, and updating the worst-case matrix by solving the inner maximization over $\mathcal U$. The method maintains an active set $\mathcal A_k$ of adversarial correlation matrices encountered so far, and at each outer iteration expands this set if a more adversarial $C$ is found. The method is based on alternatively updating $L$ and $C$. Concretely, the algorithm proceeds as follows: Initialize the active set $\mathcal A_0 = \{ \bar C \}$. Then, at each outer iteration $k = 1, \dots, K$ we have three steps:

\begin{itemize}
    \item \emph{Step 1: Laplacian update.} For the current worst case matrix $C_k = \arg \max_{C \in \mathcal A_k} \lambda_{\max}(X_{k, T} \odot C)$, we run an inner loop of $T$ steps. For each step $t = 1, \dots, T$ we sample a batch of edges $B_t \subseteq E$ with positive weight and we find a pair of edges $e^+_t$ and $e^-_t$ such that $e^+_t$ has the highest surrogate leverage $\tilde \beta_e^{\mathrm {rob}}$ in $B_t$ and $e^-_t$ has the lowest surrogate leverage in $B_t$. Then, if $\eta_t$ is the step size at step $t$, we form $\eta_t' = \min \{ \eta_t, \bar w - w_{e^+_t}, w_{e^-_t} \}$ and then we increase the weight of $e_t^+$ by $\eta_t'$ and decrease the weight of $e_t^-$ by $\eta_t'$. This yields a first order change in the surrogate disparity $\Delta \tilde g \approx - (\tilde \beta_{e_t^+}^{\mathrm{rob}} - \tilde \beta_{e_t^-}^{\mathrm{rob}})$, and Laplacian $$L_{k, t + 1} = L_{k, t} + \eta_t' \frac {m} {|B_t|} \left [ b_{e_t^+} b_{e_t^+}^T - b_{e_t^-} b_{e_t^-}^T \right ], \quad \text{s.t.} \quad \tr(L_{k, t + 1}) = 2m.$$ 

    Moreover, $X_{k, t} = (I + L_{k, t})^{-1}$ can be updated using two rank-1 updates for $e_t^-$ and $e_t^+$ respectively.
    
    \item \emph{Step 2: Group structure update.} Then, we solve the inner maximization problem by projected gradient ascent over the uncertainty set to obtain the most-violated group structure $\tilde C_k \in \mathcal U$, that is $\tilde C_k = \arg \max_{C \in \cU} \lambda_{\max}(X_{k, T} \odot C) $, given the structure $L_{k, T}$ and $X_{k, T}$. If the newly calculated objective $\lambda_{\max}( X_{k, T} \odot \tilde C_k)$ obtained by $\tilde C_k$ has larger value than the worst-case objective in the active set $\mathcal A_k$, we update the active set to include $\tilde C_k$, that is $\mathcal A_{k + 1} = \mathcal A_k \cup \{ \tilde C_k \}$. Otherwise, we set $\mathcal A_{k + 1} = \mathcal A_k$. 
\end{itemize}

\noindent Finally, we output the worst-case structure average value $\overline L_{K, T} = \frac 1 T \sum_{t = 1}^T L_{K, t}$. In the following theorem, we establish a unified approximation guarantee for the robust recommendation reweighing algorithm that bounds its output quality relative to 
to the global optimum of \eqref{eq:r_link}, assuming that we can approximately\footnote{We note that the theoretical guarantee in \cref{theorem:link_prediction_approximation} assumes the inner group structure maximization over the uncertainty set $\mathcal{U}(\rho)$ can be solved to an additive accuracy of $\varepsilon_C$. Solving the maximization problem for a general $\rho > 0$ constitutes an NP-hard, non-convex maximization over a polytope, providing a strict polynomial-time bound for $\varepsilon_C$ is computationally intractable. In practice, the framework uses projected gradient ascent warm-started at the previous estimate. While this guarantees convergence to a local maximum, $\varepsilon_C$ should be interpreted as an empirical local bound rather than a strict global guarantee.} solve the inner correlation maximization up to additive accuracy $\varepsilon_C \ge 0$: 

\begin{theorem}
\label{theorem:link_prediction_approximation}

Let $\overline L_{K,T}$ be the output of the above algorithm and let $T = \left \lceil  {\bar w^2 | E|} /{\varepsilon^2} \right \rceil$ inner reweighing steps with step size $\eta_t = \bar w \sqrt {\tfrac {|E|}  {2t}} $. Then, $\overline L_{K,T}$ satisfies

$$\max_{C \in \mathcal{U}(\rho)} h \left (\overline L_{K, T}, C \right ) \le \opt +\mu_{\max} + O(\varepsilon) + \varepsilon_C,$$

where $\opt$ is the true value of the worst case disparity, and $\mu_{\max}$ is the bound between the surrogate disparity and the true disparity (\cref{prop:surrogate_upper_bound}).

\end{theorem}

\xpar{Practical implementation} To make the algorithm scalable in real-world networks, we utilize the Johnson-Lindenstrauss lemma to approximate the leverage scores up to an $\varepsilon$ multiplicative factor using a sketch matrix $U$, by using $q = \left \lceil \log n / \varepsilon^2 \right \rceil $ i.i.d. Rademacher sketch vectors, such that $(I + L)U = R$ where $R$ is an $n \times q$ matrix containing the Rademacher vectors. Then, the robust surrogate leverage scores can be approximated as $\hat{\beta}_e^{\mathrm{rob}} = \hat x_e^T (C^* \odot vv^T) \hat x_e$, where $\hat x_e = \hat X b_e$ with $\hat X$ being the sketch approximation of $X$. The sketch can be updated using two rank-1 updates and is recomputed every $\tau$ iterations to maintain a sufficiently good approximation, where $\tau$ is chosen to minimize the trade-off between approximation error and runtime. Additional implementation details and experiments are provided in \cref{app:link_recommendation}.

\subsection{Opinion Seeding}
\label{sec:seeding}

\begin{figure}[t]
    \centering
    \includegraphics[width=0.9\linewidth]{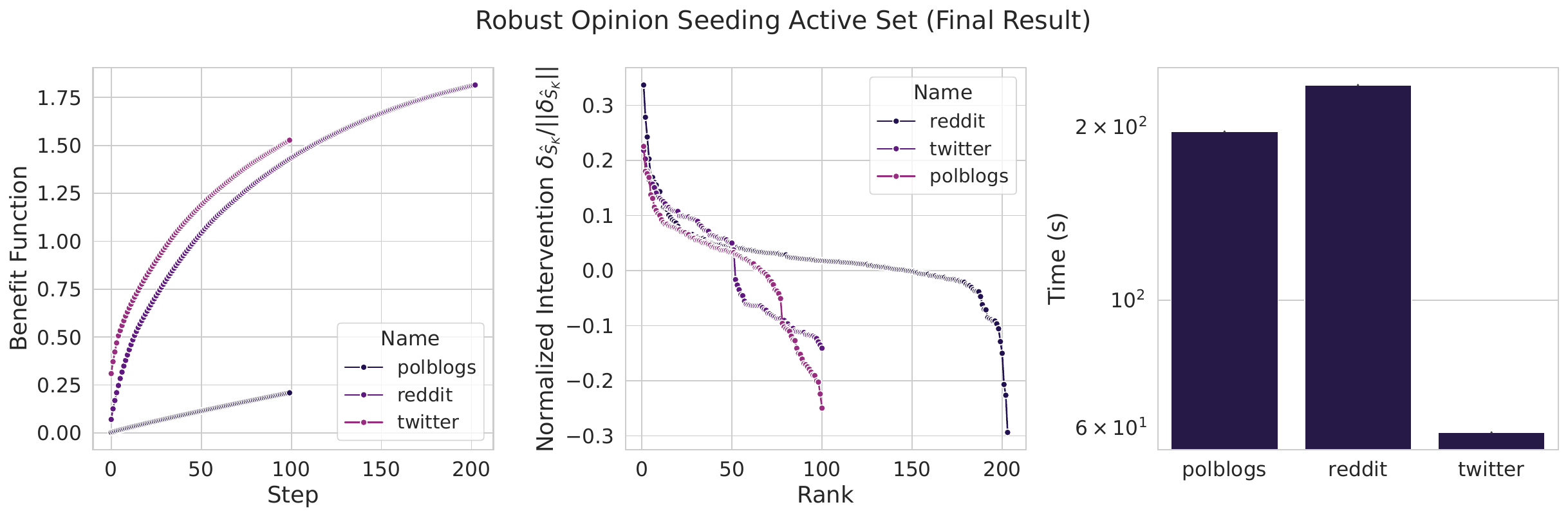}
    \caption{Seeding algorithm for $b = 100$, $\rho = 0.2$, $K = 3$, and $\nu = 2$. The final intervention $\delta_{\hat S} / \| \delta_{\hat S} \|$, which is large for a small fraction of the seeded users. \cref{sec:opinion_seeding} contains more information.}
    \label{fig:robust_seeding_main}
\end{figure}

While recommendation reweighing modifies the links, platforms may also want to use a different intervention to minimize the worst-case disparity by directly influencing a small subset of users through promoted content, verified accounts, or designated opinion leaders. Given private opinions $s$, network $L$, and uncertainty set $\mathcal{U}$, the platform selects a set $S \subseteq V$ of at most $b$ users to apply an intervention $\delta$ supported on $S$ and subject to $\| \delta \| \le 2 / \nu$ for some $\nu > 0$ to maximize worst-case disparity reduction $$\max_{S \subseteq V,  |S| \le b} \min_{C \in \mathcal{U}} F(S, L, C),$$ where $$F(S, L, C) = g(s, L, C) - \min_{\delta:  \mathrm{supp}(\delta) \subseteq S} \left \{  g(s + \delta, L, C) + \nu \| \delta \|^2 \right \}$$ is the \textit{benefit} function measuring the reduction in disparity achievable by optimally seeding
$S$ under group structure $C$. We establish in \cref{sec:opinion_seeding} that the optimal
perturbation for a fixed $S$ admits a closed form, that the
selection problem is NP-hard, and crucially that
$F(S, L, C)$ is monotone and weakly submodular in $S$ with submodularity ratio $\gamma_\nu = 1 / \kappa(Z + \nu I)$ where $\kappa(Z + \nu I)$ is the condition number of $Z + \nu I$. The weak submodularity makes the robust problem tractable: it allows the greedy hill-climbing algorithm of \citet{das2011submodular} to achieve a $(1 - e^{-\gamma_\nu})$-approximation for each fixed $C$. 

This oracle serves as the oracle inside an active set algorithm that applies a modified version of the Saturate procedure of \citet{krause2008robust} to obtain a bicriteria approximation guarantee: In short, the algorithm proceeds similarly to the recommendation reweighing algorithm by maintaining an active set $\mathcal A_k$ over $k = 1, \dots, K$ outer iterations. At each iteration $k$ the algorithm applies the modified Saturate procedure of \citet{krause2008robust} with a budget $b_{k} = \left ( 1 + \frac {\log (k \alpha / \varepsilon)} {\gamma_{\nu, k}'} \right )b$ over the active set $\mathcal A_k$ (which corresponds to the set of active scenarios) and obtains a target set $\hat S_k$ where $\gamma_{v, k}'$ is the worst case submodularity ratio of any group structure $C \in \mathcal A_k$, $\varepsilon$ is a tolerance parameter, and $\alpha$ is a lower bound to the robust optimum. During the second step, the algorithm finds the worst-case group structure $\hat C_k$ that minimizes $F(\hat S_k, L, C)$ over $\mathcal U$. If $F(\hat S_k, L, \hat C_k) < \min_{C \in \mathcal A_k} F(\hat S_k, L, C)$ then we add $\hat C_k$ to the active set, i.e., $\mathcal A_{k + 1} = \mathcal A_k \cup \{ \hat C_k \}$. Otherwise, the active set remains the same, i.e., $\mathcal A_{k + 1} = \mathcal A_k$. Finally, the algorithm outputs the solution $\hat S_K$ of the last iteration. 

Similarly to \cref{sec:link_recommendation}, given an oracle that finds the worst-case group structure up to accuracy $\varepsilon_C$, the robust opinion seeding returns a final set $\hat S_K$ such that $\min_{C \in \mathcal U} F(\hat S_K, L, \hat C) \ge \alpha - (\varepsilon + \varepsilon_C),$ where $\alpha \in [0, \opt]$, and $\opt =  \max_{S : |S| \le b}  \min_{C \in \mathcal U} F(S, L, C)$ is the robust optimum. Due to length constraints, we defer the full analysis of the algorithm to \cref{sec:opinion_seeding}.



\section{Experiments} \label{sec:experiments}


We conduct experiments on three differing real-world social networks: \textit{Polblogs}~\citep{adamic2005political}, \textit{Reddit}, and \textit{Twitter} from \citet{Chitra2020}. We compare against baselines, examine the impact of our algorithms on the platform microstructure, and consider a realistic scenario in which the platform trains a predictive model to infer membership.

\subsection{Recommendation Reweighing} \label{sec:experiments_link_recommendation}

\xpar{Comparison with baselines and impact on the network structure} For each network, the platform computes a nominal estimate $\bar C = \bar r \bar r^T$ with error margin $\rho \in [0, 1/2]$, which forms the basis of the uncertainty set $\mathcal U(\rho)$; the nominal membership is obtained by spectral clustering, $\bar r = \sign(v_2)$, where $v_2$ is the Fiedler eigenvector of the Laplacian. We set $\varepsilon = 0.1$ and choose the iterations ($T, K$) and step size ($\eta_t$) according to \cref{theorem:link_prediction_approximation}. Using subscripts ``$\mathrm{old}$'' and ``$\mathrm{new}$'' for input and output quantities, we report the percentage change between the structural disparity and its surrogate before intervention, $\lambda_{\max}(M_{\mathrm{old}} \odot \bar C)$ and $\lambda_{\max}(X_{\mathrm{old}} \odot \bar C)$, and after the intervention, $\lambda_{\max}(M_{\mathrm{new}} \odot C_{\mathrm{new}})$ and $\lambda_{\max}(X_{\mathrm{new}} \odot C_{\mathrm{new}})$\footnote{We note that even if the surrogate disparity is an upper bound to the true disparity, that is not be the case when measuring the percentage change as the ordering is not preserved.}.



\begin{figure}[t]
    \centering
    \includegraphics[width=0.95\linewidth]{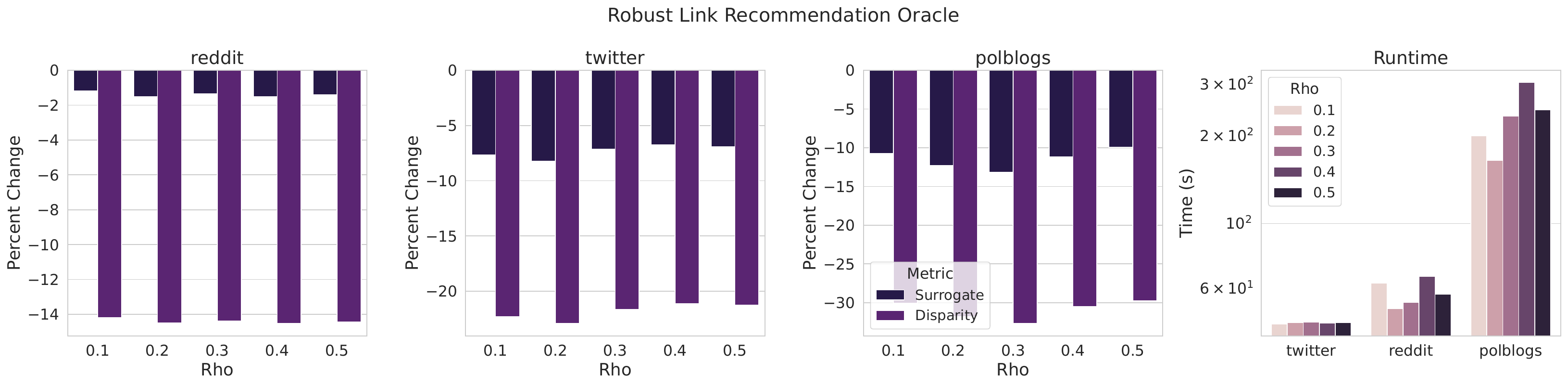}\\
    \includegraphics[width=0.95\linewidth]{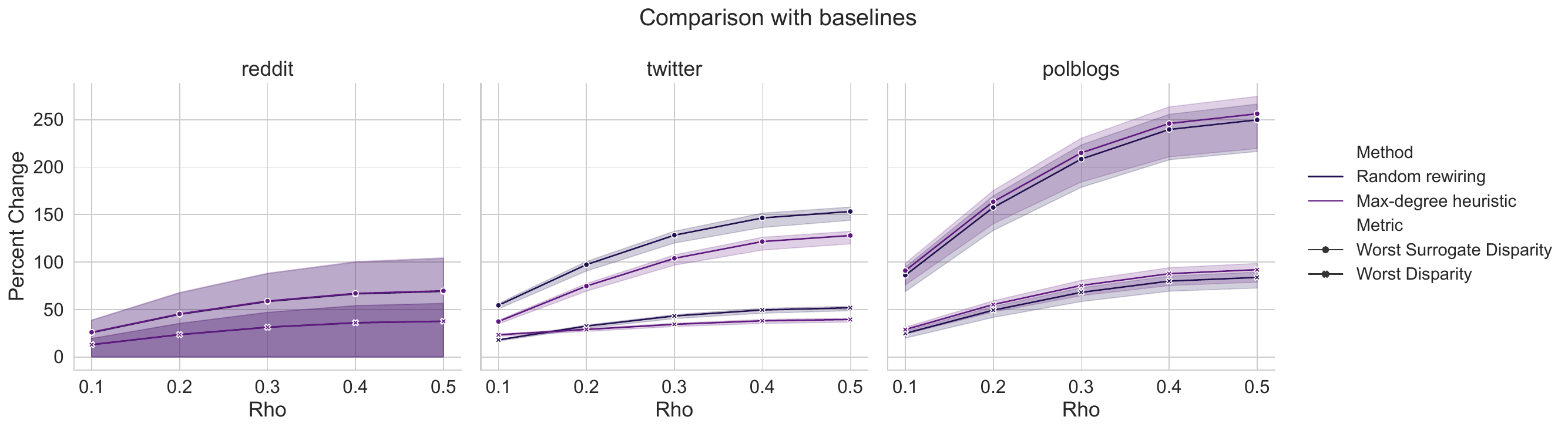}
    \caption{Top: Percent changes of the disparity, the surrogate, and the polarization as a result of running the robust recommendation reweighing oracle for various values of $\rho$. Additionally, we report the runtime for each network. The runtime ranges from $10^1-10^2$ seconds. Bottom: Comparison with baselines (random rewiring and max-degree heuristic)}
    \label{fig:robust}
\end{figure}

We run the robust recommendation reweighing algorithm for various values of $\rho$ and report the percent change in the worst-case objective, the surrogate, and the polarization. The runtime is in the order of $10^1$-$10^2$ seconds across the three networks, showing that our method scales to large networks.

\noindent\textbf{Performance of the robust oracle.}
\cref{fig:robust} reports the percent change in the worst-case disparity $g$, surrogate disparity $\tilde{g}$, after running the robust recommendation reweighing algorithm across values of $\rho$. The percent changes have mixed signs across datasets (mostly positive) and values of $\rho$, with some settings exhibiting modest reductions and others exhibiting modest increases.
 
This behavior is a direct consequence of the robust formulation, which optimizes the surrogate $\tilde g$ against the \emph{worst-case} $C \in \mathcal{U}(\rho)$ rather than the nominal $\bar{C}$. \cref{fig:robust} (bottom row) reports the same metrics for two uninformed baselines (random rewiring and a max-degree heuristic) that reweight edges without regard for the group correlation structure. Both produce substantially larger positive percent changes, often exceeding $100\%$, versus $1\%$--$10\%$ for our algorithm, indicating that uninformed interventions severely worsen the platform's worst-case exposure to group-level consensus sensitivity. Thus, our robust algorithm avoids deterioration of the worst-case objective across a wide range of $\rho$.
 
This has an important implication: a platform that optimizes for the nominal group structure alone, or that applies heuristic reweighting, can inadvertently make the consensus far more vulnerable to adversarial shifts in group composition, even if nominal performance appears acceptable. The robust formulation instead provides a safety guarantee at the cost of a bounded reduction in nominal performance.

\begin{figure}
    \centering
    \includegraphics[width=0.98\linewidth]{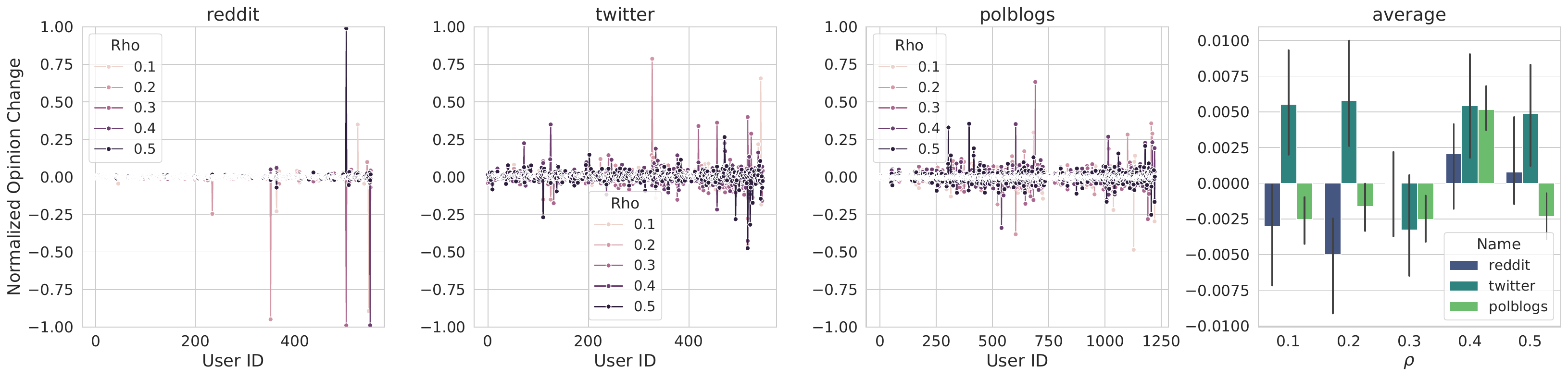}\\
    \includegraphics[width=0.85\linewidth]{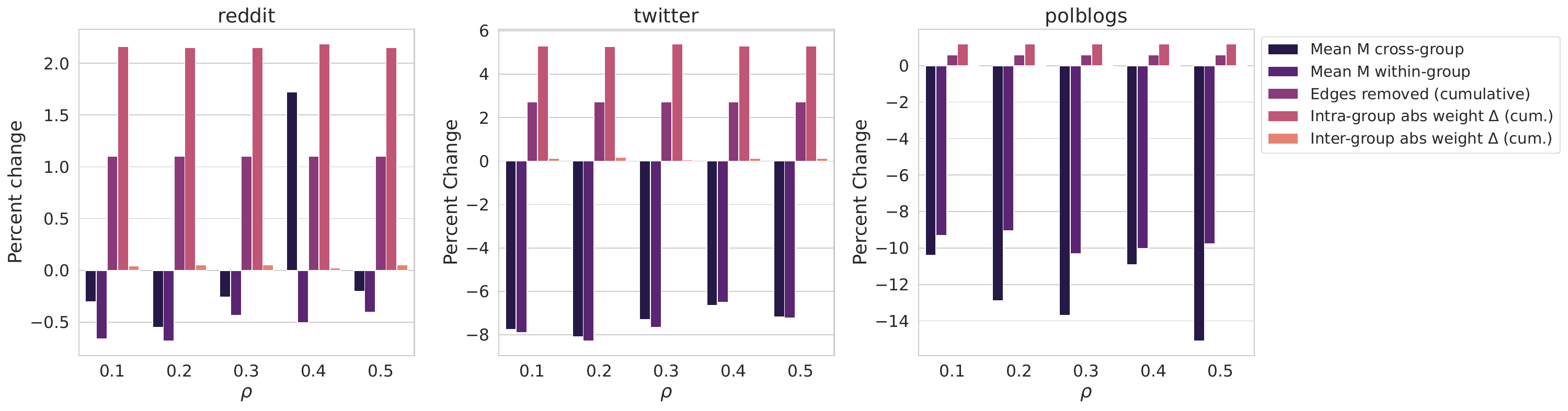}
    \includegraphics[width=0.85\linewidth]{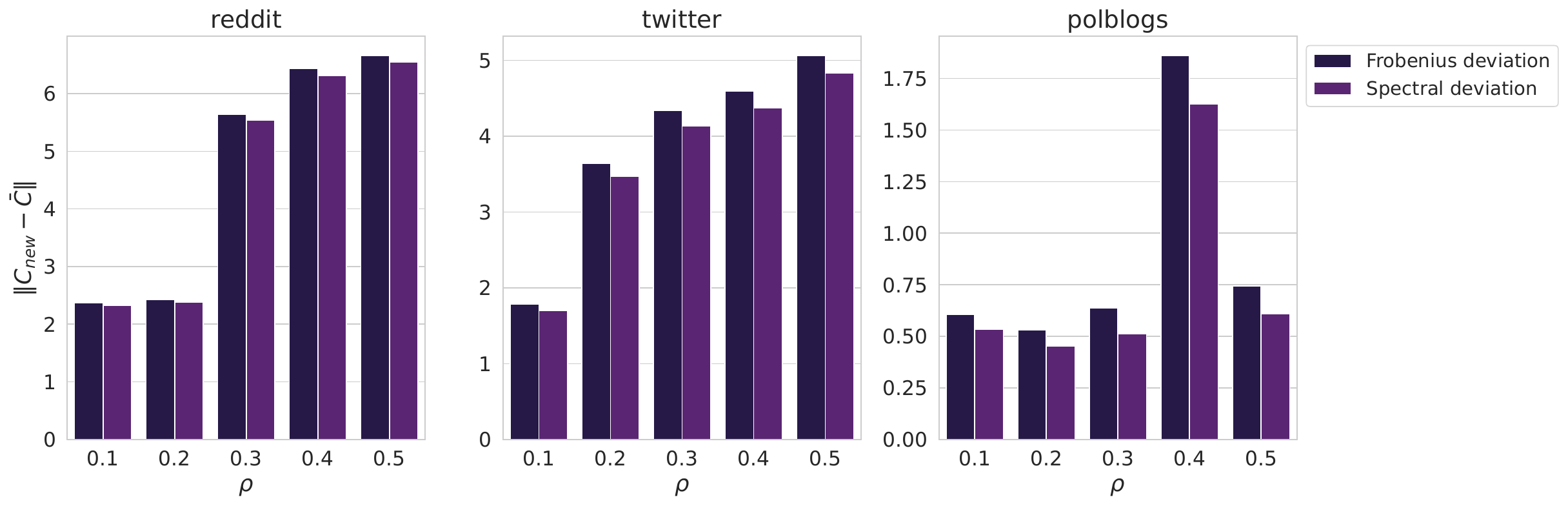}
    \caption{Top: Normalized opinion change per user after the robust recommendation reweighing algorithm as a function of the classifier error $\rho$, and average among all users. We do not observe significant changes from 0 in most cases, which shows that our robust recommendation reweighing algorithm is stable and does not significantly mutate the opinions. Bottom left: Impact on the links of the network as a function of $\rho$. Specifically, we measure the mean value of $M_{\mathrm{new}} = (I + L_{\mathrm{new}})^{-1} \odot C_{\mathrm{new}}$ cross-group ($\bar r_i \neq \bar r_j$) and within-group ($\bar r_i = \bar r_j$), as well as the intervention variation (cumulative absolute value of the change in the weights), the number of edges removed during the algorithm's execution, the intra-group absolute weight, and the inter-group absolute weight. We observe significant differences for different values of $\rho$ and the dataset. Bottom right: We report the percent change in the final group structure $C_K$ against the nominal estimate $\bar C$ in terms of the Frobenius norm ($\| C_{\mathrm{new}} - \bar C \|_F$) and the spectral norm ($\| C_{\mathrm{new}} - \bar C \|$) as a function of $\rho$. We observe significant differences across datasets and values of $\rho$, with more substantial changes corresponding to higher values of $\rho$.}
    \label{fig:impact}
\end{figure}

\noindent\textbf{Impact on the platform microstructure.}
We also study the algorithm's impact on the network structure. The top panel of \cref{fig:impact} shows the normalized opinion change $\frac {z_{\mathrm{new}} - z_{\mathrm{old}}} {\| z_{\mathrm{new}} - z_{\mathrm{old}} \|}$ per user after running the robust algorithm. Across all three networks and all values of $\rho$, the per-user changes are small and centered near zero with no systematic drift, and the average change across users (rightmost panel) is consistently close to zero, confirming that the algorithm achieves its objective through structural modification. This is a desirable governance property: the algorithm redistributes influence through the interaction structure without significantly suppressing or amplifying any individual user's voice. Next, the bottom-left panel of \cref{fig:impact} decomposes the impact on the link structure. In accordance to the opinion changes, we measure the intervention's impact on the links by measuring the mean value of $M_{\mathrm{new}} = (I + L_{\mathrm{new}})^{-1} \odot C_{\mathrm{new}}$ regarding cross-group ($\bar{r}_i \neq \bar{r}_j$) and within-group ($\bar{r}_i = \bar{r}_j$) components. We observe that the two measures exhibit a small percent change, showing that the algorithm can achieve minimization of the maximum disparity without incurring significant changes to the network structure, even if the worst-case group structure $C_{\mathrm{new}}$ is significantly different from the nominal group structure $C_{\mathrm{old}} = \bar C$.

Finally, the bottom-right panel of \cref{fig:impact} tracks the Frobenius and spectral norm deviation of the final adversarial matrix $C_{\mathrm{new}}$ from the nominal estimate $C_{\mathrm{old}} = \bar{C}$; both grow monotonically with $\rho$. Thus, even though the platform's belief about the worst-case correlation structure can change substantially, the algorithm minimizes the worst-case disparity with minimal changes to the network and user opinions.

\begin{figure}
    \centering
    \includegraphics[width=0.92\linewidth]{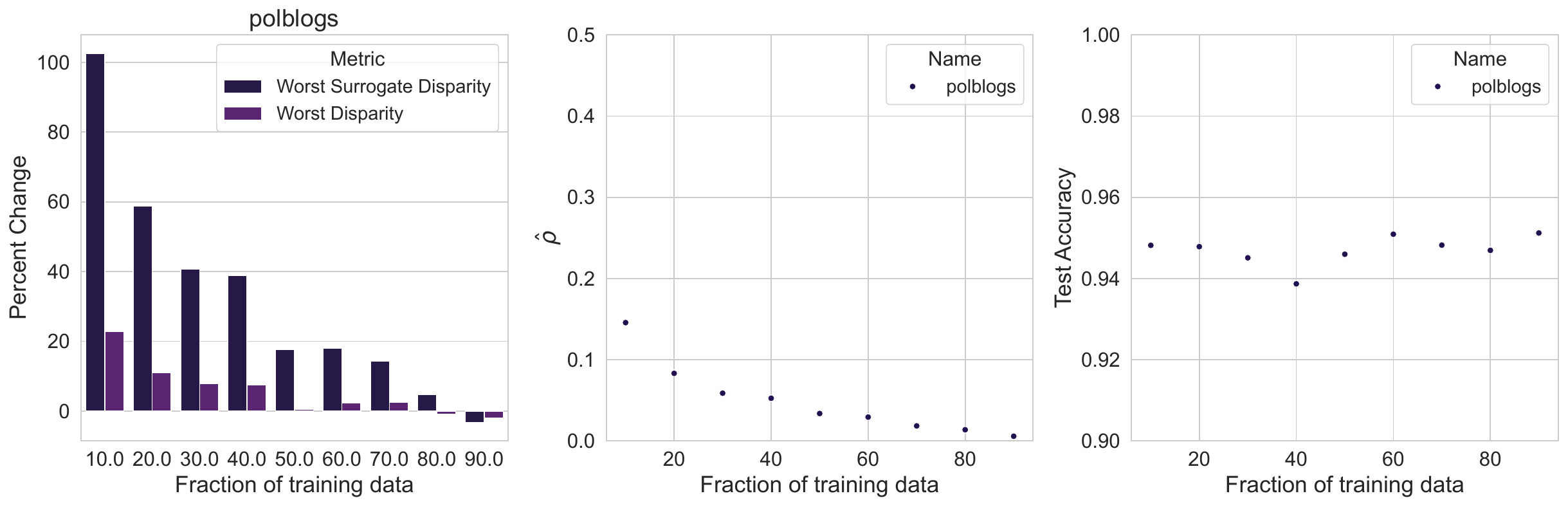}
    \caption{Percent change in the metrics when a logistic regression predictive model is employed by the platform for the Polblogs dataset. The logistic regression model is trained on $\lceil p n \rceil$ datapoints using user embeddings for values of $p$ ranging from $10\%$ of the users to $90\%$ of the users. The figure reports the change in the metrics as a function of the amount of training data available.}
    \label{fig:predictive_model}
\end{figure}

\xpar{Practical implementation using a predictive model} The previous experiments assumed the platform knows its classifier error $\rho$  exactly. Here, we come back to the managerial example we motivated in \cref{sec:managerial_example} and test a more realistic setting for the platform: the platform learns group membership from a real classifier trained on limited labeled data, and the uncertainty set is estimated from the classifier's validation error.

\cref{fig:predictive_model} examines a realistic deployment scenario in which the platform has user features $\{ x_i \}_{i \in [n]}$ for all users and binary labels $\{ y_i \}_{i \in V_0}$ for a subset $V_0$ of size $pn$. The platform trains a model $\hat f$ on the Node2Vec embeddings \citep{grover2016node2vec} to predict all users' labels and constructs a noisy estimate $\bar C$: for $i \in V_0$ we set $\bar r_i = 2y_i - 1$, and otherwise $\bar r_i = 2 \one \{ \hat f(x_i) \ge 0.5 \} - 1$. We use a high-probability estimate $\hat \rho = (1 - p)( \hat \rho_{\mathrm{val}} + \hat \sigma_{\mathrm{val}})$, where $\hat \rho_{\mathrm{val}}$ and $\hat \sigma_{\mathrm{val}}$ are the mean and standard deviation of the validation accuracy after five-fold cross-validation. The ground-truth labels for $V_0$ are the original political-affiliation labels from \citep{adamic2005political}.

We observe that as the fraction $p$ of the training data increases, the error estimate $\hat \rho$ decreases and the uncertainty set $\mathcal U(\hat \rho)$ becomes smaller. More training data yields a more accurate classifier, which tightens the uncertainty set and allows the algorithm to optimize more precisely, resulting in a lower worst-case disparity change under the robust algorithm. 

\subsection{Opinion Seeding}

We validate the opinion seeding algorithm on the same three real-world networks using a nominal seeding budget $b = 100$, $\rho = 0.2$, $K = 3$ outer iterations of the robust algorithm (see \cref{fig:robust_seeding} in \cref{sec:opinion_seeding}). Due to length constraints, full experimental details, pseudocode, and proofs have been deferred to \cref{sec:opinion_seeding}. Our algorithm outperforms all three uninformed baselines, random selection, max-degree selection, and PageRank-based selection, often by a substantial margin. The gap is most pronounced on Polblogs, where the community structure is most clearly separated, and the alignment between the greedy seed set and the group boundary is the largest; max-degree and PageRank heuristics perform comparably to each other but remain well below the greedy solution, confirming that selecting high-influence users without regard for disparity reduction is insufficient. Additionally, the resulting interventions $\delta_{\hat S}$ are small across the majority ($\ge 90\%$) of seeded users and datasets, which consistently shows, together with the results of \cref{sec:experiments_link_recommendation}, that our robust formulations do not need to significantly alter the platform microstructure in order to be effective, which makes the suggested platform's policy more auditable. 

\section{Discussion} \label{sec:discussion}

The results above carry direct implications for the platform and can be directly operationalized.

\subsection{Managerial Insights}

\begin{figure}[t]
    \centering
    \includegraphics[width=0.55\linewidth]{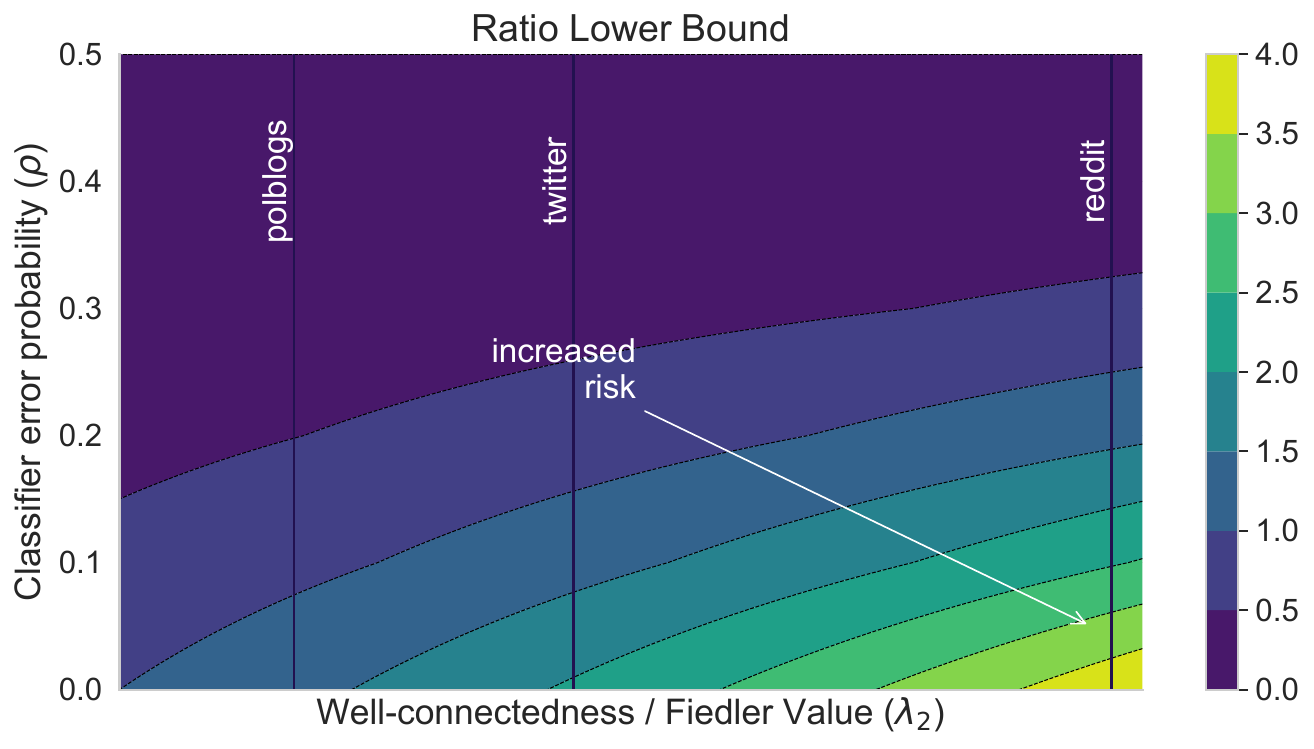}
    \caption{Illustration of the ratio lower bound from \cref{theorem:gap}. If the well-connectedness ($\lambda_2$) of the network is high, and the predictive model error ($\rho$) is low, then the network is more vulnerable. Vertical lines that correspond to each network and its well-connectedness are drawn. Regarding the datasets used in \cref{sec:experiments}, the Reddit network is in the high-risk region.}
    \label{fig:vulnerability}
\end{figure}

\xpar{Disparity as a complementary metric to polarization} A platform that measures polarization only, which corresponds to some measure of variance of the public opinions (or equivalently sentiment), can be severely misled about its exposure to group-level consensus risk. Our managerial example in \cref{sec:managerial_example} illustrates this: the Polblogs network exhibits a polarization value of 0.1, yet simultaneously achieves a significantly higher disparity, meaning the consensus outcome is effectively determined entirely by one ideological cluster. As \cref{theorem:gap} states, this gap is not an artifact of an unusual network; it arises whenever the network is well-connected, and the error is low, as shown in \cref{fig:vulnerability}. 

The practical implication is direct: disproportionate group influence should be tracked alongside polarization. Disparity captures a distinct dimension of systemic risk, namely, whose opinions structurally drive the outcome and how sensitive that outcome is to shifts in group-level participation. Monitoring both metrics together provides early warning of risks that polarization alone cannot detect, such as adversarial manipulation, single-cluster amplification, or the gradual erosion of cross-group influence. Importantly, computing disparity requires only partial knowledge of the group structure, obtainable by training a predictive model with limited data.


\xpar{Heuristics are not efficient for reducing disparity} A platform that attempts to reduce disparity using uninformed heuristics, randomly promoting cross-group connections, or even reweighting edges by sensible heuristics such as the popularity of accounts, risks causing disparity to \emph{increase} substantially. Concretely, we find that across the Reddit, Twitter, and Polblogs datasets, random and degree-based baselines produce worst-case disparity changes approximately 10 times larger than those of our robust algorithm, and in the most severe cases drive surrogate disparity up by several hundred percent. The reason is structural: which edges to reweight depends sensitively on the group partition, and an edge that reduces disparity under the true partition can be harmful under a less specified one. Because group membership can be estimated from noisy behavioral proxies and imperfect classifiers, a heuristic that ignores this uncertainty may systematically target the wrong edges.
 
In practice, platforms exercise these interventions through feed-ranking, content-exposure policies, and recommendation engines: the weight $w_{ij}$ corresponds to the effective probability that user $i$ is exposed to and influenced by user $j$, so adjusting the ranking of $j$'s content in $i$'s feed is equivalent to adjusting $w_{ij}$. Our robust reweighing algorithm outputs per-edge leverage scores that rank which pairwise interactions, if reweighted, would most reduce worst-case disparity; a platform can apply these to feed-ranking parameters, up-weighting cross-group exposure for high-leverage pairs and reducing intra-group amplification for those that concentrate influence. As \cref{fig:impact} confirms, these interventions are stable and alter the microstructure minimally, which is desirable from a governance standpoint.


\xpar{Our robust algorithms capture realistic and dynamical scenarios} Platforms do not intervene with perfect knowledge of group membership, and the groups themselves are not static: they shift as users join, leave, and change behavior, and are vulnerable to the same coordinated manipulation that causes systemic fragility. The uncertainty set models this directly, and the robust algorithm produces interventions that perform well not for a single estimated group structure but against the worst case within a plausible neighborhood; this trade-off can be controlled by enlarging the uncertainty set. Our evaluation shows that robust interventions have a bounded impact on both the network structure and the users' expressed opinions (see \cref{fig:impact}).

\xpar{Implications for platform governance} Several jurisdictions are developing algorithmic-auditing requirements, including obligations to assess whether recommendation systems systematically amplify certain viewpoints at the expense of others. Consensus disparity provides a precise, computable operationalization of this desideratum: a high disparity value is quantitative evidence that the consensus outcome is disproportionately driven by one group's private opinions, regardless of the groups' relative sizes. Platforms can use it to generate periodic auditing reports stratified by community type and track its joint trajectory with polarization: low polarization with high disparity signals that low opinion variance is driven by one group's dominance rather than convergence, while rising polarization with declining disparity reflects a balanced diversification of influence.

The intervention algorithms are also directly auditable: The robust reweighing algorithm produces an auditable ranked list of targeted edge adjustments that is dependent on the platform's beliefs about the group structure; the seeding algorithm produces an auditable list of targeted accounts, intervention directions, and magnitudes. Platforms that adopt these tools can build an evidentiary record demonstrating that their interventions are principled, calibrated to uncertainty, and non-discriminatory; a significant advantage in cases of auditing.

\subsection{Extensions of the Framework and Future Work}
\label{sec:extensions}
 
We have presented the disparity metric for the canonical case of a single binary partition under the FJ model, as this is the setting that admits clean characterizations and is connected to the existing literature. The framework, however, is considerably more general than this presentation suggests, and three extensions of direct managerial relevance follow with little additional machinery. We sketch them here at a high level and develop them formally in Appendix~\ref{app:generalizations}.
  
First, many users may belong to more than one group, and their membership in each group is not necessarily binary. The disparity metric we developed naturally extends to more than two groups, and group membership may not be binary, either by aggregating the pairwise disparities into a single index, by tracking the worst-case over group pairs, or by reporting a compact matrix-valued index of how influence is distributed across all groups. The robust intervention algorithms of \cref{sec:interventions} can generalize in these settings as well, allowing a platform to audit and mitigate group-level consensus risk across a full set of multiple and possibly overlapping communities.
 
Second, while we instantiate the metric on the FJ model, the construction depends on the consensus dynamics only through their being linear: any model of the form $z=Ts$ that aggregates opinions through a
linear operator $T$ admits a disparity metric of the same form. This covers common variants such as users with varying susceptibility to persuasion (see \citet{Abebe2018} and the DeGroot model \citep{DeGrootModel}, so the diagnosis and the interventions are not tied to one particular model of social influence.
 
Finally, several directions can extend naturally as future work. First, the framework currently treats the
network as undirected; extending to directed graphs would capture asymmetric influence relationships that are common in follower-based social platforms. Second, the opinion seeding problem treats the seed set as the platform's only targeted intervention; combining recommendation reweighing and seeding into a joint intervention problem may yield stronger guarantees. Third, while we model group membership uncertainty through a classifier error model, richer
uncertainty sets, for instance, those derived from differential privacy constraints on label collection, may be tractable within the same robust framework and would connect directly to platform data governance practices. 


\ACKNOWLEDGMENT{Supported in part by a Simons Investigator Award, a Vannevar Bush Faculty Fellowship, AFOSR grant FA9550-19-1-0183, a Simons Collaboration grant, and a grant from the MacArthur Foundation. The authors would like to thank Panagiotis Adamopoulos, Olivia Liu Sheng, and M. Amin Rahimian for their feedback and suggestions on the paper.}

{\SingleSpacedXII
\bibliographystyle{informs2014} 
\bibliography{references} 
\par}



\newpage

\ECSwitch
\begin{center}
\ECHead{Online Companion to ``Quantifying and Mitigating Consensus Disparity in Social and Information Networks''}
{\large Marios Papachristou, Jon Kleinberg}
\end{center}

\crefalias{section}{appendix}
\crefalias{subsection}{appendix}
\crefalias{subsubsection}{appendix}

\makeatletter

\let\EC@origaddcontentsline\addcontentsline

\renewcommand{\addcontentsline}[3]{%
  \def\EC@type{#2}%
  \def\EC@section{section}%
  \ifx\EC@type\EC@section
    \EC@origaddcontentsline{ectoc}{#2}{#3}%
  \fi
}

\newcommand{\ECtableofcontents}{%
  \section*{E-Companion Table of Contents}
\begingroup
  \hypersetup{linktoc=none}

  \setcounter{tocdepth}{1}%
  \renewcommand*\l@section{\@dottedtocline{1}{0em}{3em}}%

  \makeatletter
  \renewcommand*\numberline[1]{%
    \hb@xt@\@tempdima{\hyperlink{\Hy@tocdestname}{##1}\hfil}%
  }%
  \makeatother

  \@starttoc{ectoc}%
\endgroup
}

\makeatother




%
%


\makeatletter
\renewcommand{\p@subfigure}{\thefigure}      
\renewcommand{\thesubfigure}{(\alph{subfigure})}
\makeatother

\ECtableofcontents

\medskip

\noindent\textbf{Organization.} \cref{app:definition} formalizes the disparity metric, deriving the zero-baseline attribution map from three axioms and relating it to a Shapley decomposition of the consensus. \cref{app:link_recommendation} gives implementation details and additional experiments for the recommendation-reweighing intervention, and \cref{sec:opinion_seeding} develops the opinion-seeding intervention, including the optimal perturbation, NP-hardness, weak submodularity, the robust active-set algorithm, and its experiments. \cref{app:proofs} collects the proofs of all results, \cref{app:helper_lemmas} the supporting lemmas. Finally, \cref{app:generalizations} extends the framework.

\begin{APPENDIX}{}

\section{Definition and Properties of the Disparity} \label{app:definition}

In platform-governance settings the opinions of the silenced group are unobserved, suppressed, or absent, so a disparity measure sensitive to how those absent opinions are parametrized would capture an artifact rather than a structural property; we therefore require invariance to affine reparametrizations of the silenced group. Fix $G=(V,E)$ with Laplacian $L$ and $\|s\|=1$. Given a partition $(A,\bar A)$, an \emph{attribution map} is a function $\Phi:\mathbb R^n\times 2^V\to\mathbb R^n$ in which $\Phi(s,A)$ is the consensus attributable to $A$; let $s^{(\alpha,\beta)}$ replace $s_i$ by $\alpha s_i + \beta$ for $i\in\bar A$ while keeping $s_i$ for $i\in A$. We require three properties: \emph{(i) additivity}, $\Phi(s,A)+\Phi(s,\bar A) = (I+L)^{-1}s$; \emph{(ii) affine invariance}, $\Phi(s^{(\alpha,\beta)},A) = \Phi(s,A)$ for all $\alpha,\beta\in\mathbb R$ (equivalently, $\Phi(s,A)$ depends on $s$ only through $s_A = s\odot\one_A$); and \emph{(iii) linearity} of $s\mapsto\Phi(s,A)$ for each fixed $A$.

\begin{theorem}
\label{thm:affine_invariance}
The zero-baseline attribution map $\Phi(s,A) = (I+L)^{-1}s_A$ is the unique map satisfying additivity, affine invariance, and linearity.
\end{theorem}

\xpar{Proof} By linearity, $\Phi(s,A) = M_A s$ for some $M_A \in \mathbb R^{n\times n}$. Affine invariance with linearity gives $M_A\big[(\alpha-1)s_{\bar A} + \beta\one_{\bar A}\big] = 0$ for all $s$ and all $\alpha,\beta$, where $s_{\bar A} = s\odot\one_{\bar A}$. Taking $\alpha=2,\beta=0$ yields $M_A s_{\bar A} = 0$ for all $s$, so the columns of $M_A$ indexed by $\bar A$ vanish, i.e.\ $M_A = M_A\diag(\one_A)$; the same argument applied to $\bar A$ shows the columns of $M_{\bar A}$ indexed by $A$ vanish. Combining with additivity $M_A + M_{\bar A} = (I+L)^{-1}$: for $j\in A$, $M_{\bar A}e_j = 0$ so $M_A e_j = (I+L)^{-1}e_j$, while for $j\in\bar A$, $M_A e_j = 0$. Hence $M_A = (I+L)^{-1}\diag(\one_A)$ and $\Phi(s,A) = (I+L)^{-1} s_A$. 

\subsection{Connection to the Shapley Values}
\label{app:shapley}

For a finite player set $N = \{1,\ldots,n\}$ and characteristic function $v:2^N\to\mathbb R$ with $v(\emptyset)=0$, the \emph{Shapley value} $\phi_i(v) = \sum_{S\subseteq N\setminus\{i\}} \tfrac{|S|!\,(|N|-|S|-1)!}{|N|!}\,[v(S\cup\{i\}) - v(S)]$ is the unique attribution satisfying efficiency, symmetry, the null-player property, and linearity \citep{shapley1953value}. We instantiate it for the linear opinion model $z = (I+L)^{-1}s$ through the following attribution game.

\begin{definition}
Given $G = (V,E)$ with Laplacian $L$ and $\|s\| = 1$, define the characteristic function $v_s: 2^V \to \mathbb{R}^n$ by $v_s(A) = (I+L)^{-1} s_A$ with $s_A = s \odot \one_A$, i.e.\ the consensus obtained if only users in $A$ hold their private opinions and the rest are silenced.
\end{definition}

\begin{proposition}
\label{prop:shapley-closed-form}
For the attribution game $v_s$, the Shapley value of user $i$ is $\phi_i(v_s) = (I+L)^{-1} s_i e_i$, where $e_i$ is the $i$-th standard basis vector.
\end{proposition}

\noindent \emph{Proof.} By linearity of $(I+L)^{-1}$ and $s_A = \sum_{i \in A} s_i e_i$, the marginal contribution $v_s(A \cup \{i\}) - v_s(A) = (I+L)^{-1} s_i e_i$ is independent of $A$, so its weighted average over $A \subseteq V \setminus \{i\}$ equals the same vector.

Consequently the group attributions are $z_A = \sum_{i \in A} \phi_i(v_s)$ and $z_{\bar A} = \sum_{i \in \bar A} \phi_i(v_s)$ with $z_A + z_{\bar A} = z$, so the conditional disparity $f(s,L,A) = \|z_A - z_{\bar A}\|^2$ is exactly the squared norm of the difference in Shapley attributions between the two groups: a large value means one group's contributions dominate the consensus. Taking the expectation over the partition gives the average disparity
\begin{align}
    g(s, L, C) = \ev {(A, \bar A) \sim C} {\Big\| \sum_{i \in A} \phi_i(v_s) -\sum_{i \in \bar{A}} \phi_i(v_s) \Big\|^2}.
\end{align}

\section{Recommendation Reweighing} \label{app:link_recommendation}

\subsection{Additional Experiments}

In \cref{fig:steps}, we present additional experiments for the two oracles, corresponding to Step 1 (top panel) and Step 2 (middle panel) described in \cref{sec:link_recommendation} respectively, as well as the percent change after finding the optimal  optimal network structure when the group structure is set as in \cref{eq:correlation_matrix} (bottom panel): 

\begin{align*}
    C_{ij}(\rho) = \begin{cases}
        1, & i = j \\
        (1 - 2 \rho)^2 \bar r_i \bar r_j, & i \neq j 
    \end{cases}
\end{align*}

\begin{figure}[!h]
    \centering
    \includegraphics[width=0.78\linewidth]{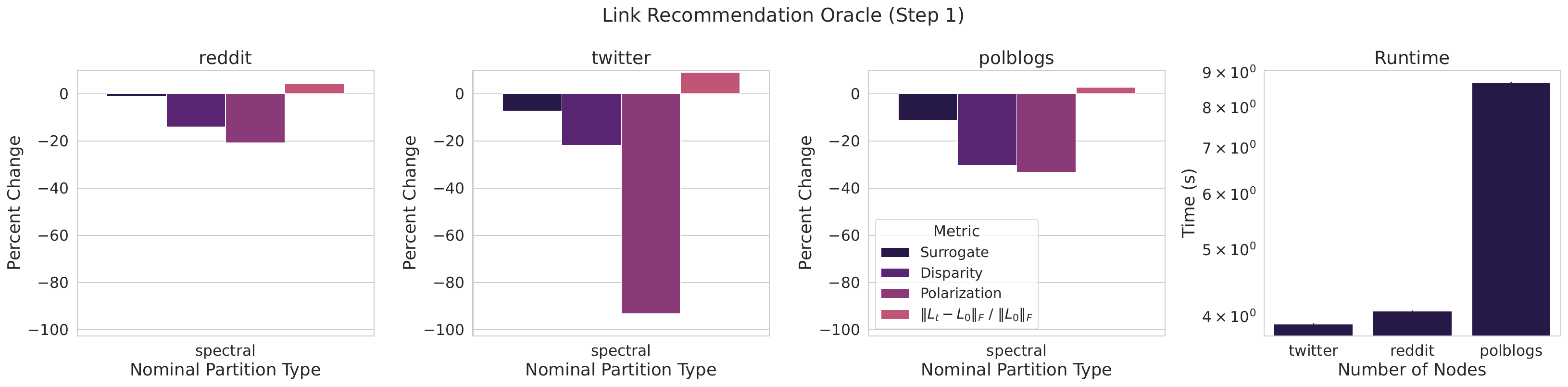}
    \includegraphics[width=0.78\linewidth]{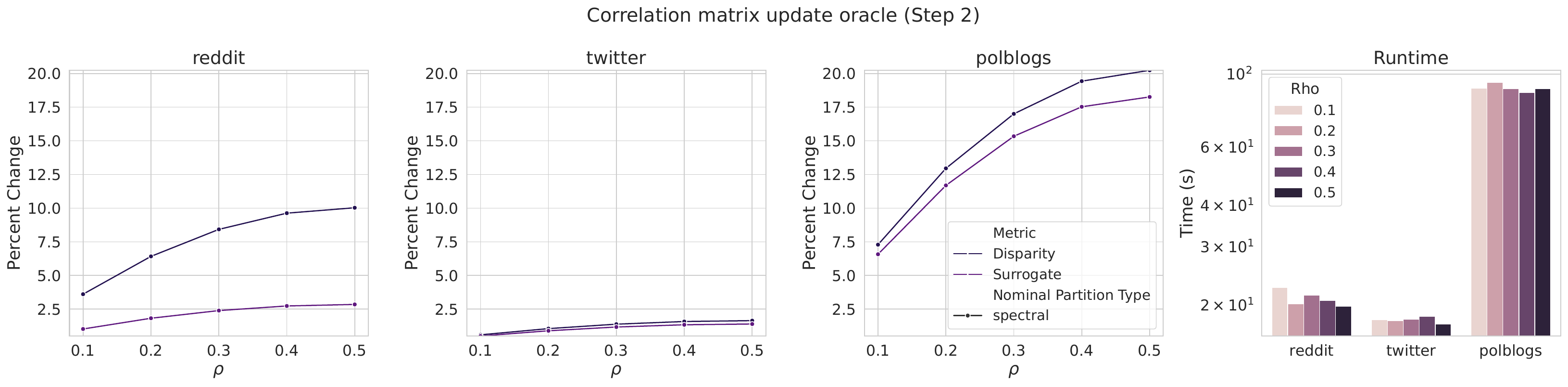}
    \includegraphics[width=0.7\linewidth]{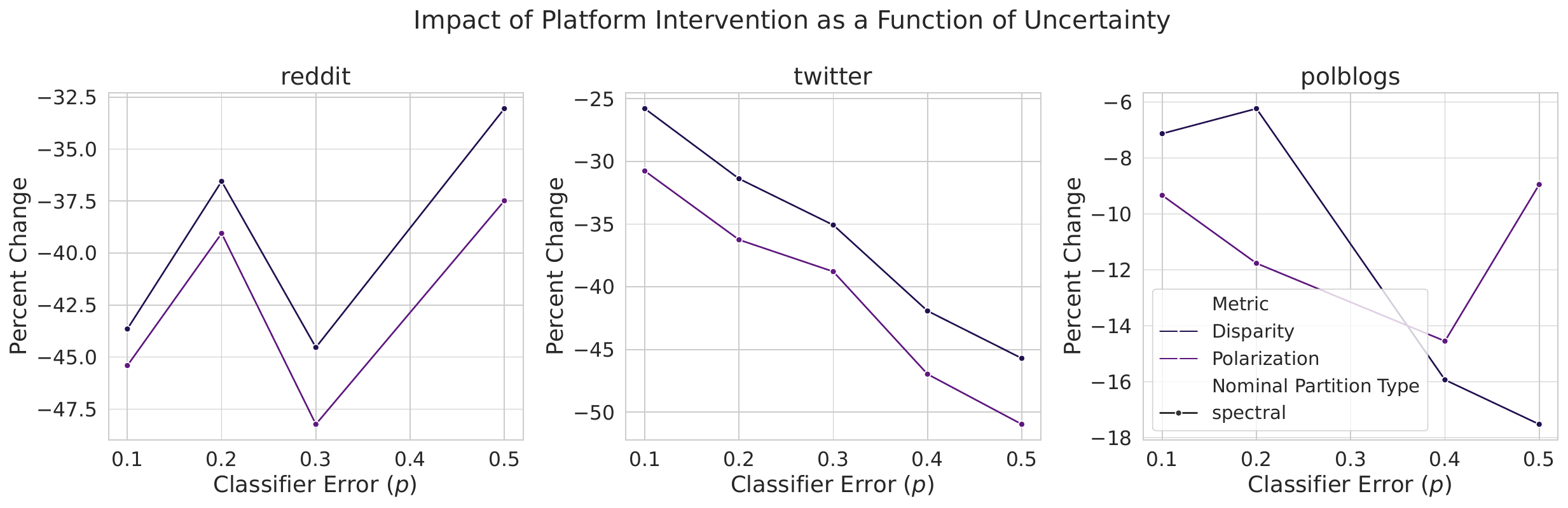}
    \caption{Top: Performance of the link recommendation oracle under the nominal group structure $\bar C = \bar r \bar r^T$ (Step 1). We report the percent change in the surrogate disparity, the true disparity, the polarization, and the network Laplacian for the three datasets. Middle: Worst-case percent change of the disparity and the surrogate by running the group structure update oracle (Step 2) on the initial network for various values of $\rho$. Bottom: Percent change in the metrics when the network is optimized (Step 1), assuming that the group structure is fixed and set according to \cref{eq:correlation_matrix} of \cref{sec:managerial_example} for values of $\rho \in \{ 0.1, 0.2, 0.3, 0.4, 0.5 \}$.}
    \label{fig:steps}
\end{figure}

\section{Opinion Seeding} \label{sec:opinion_seeding}

Recall the robust opinion-seeding problem from \cref{sec:seeding}: the platform selects $S \subseteq V$ with $|S| \le b$ and a perturbation $\delta$ supported on $S$ to solve
\begin{align}
    \max_{S \subseteq V,\ |S| \le b}\ \min_{C \in \mathcal{U}}\ F(S, L, C),
    \tag{R-Seed}
    \label{eq:r-seed}
\end{align}
where $F(S, L, C) = g(s, L, C) - \min_{\delta:\ \mathrm{supp}(\delta) \subseteq S} g(s + \delta, L, C)$ is the benefit function (the nominal case $\mathcal{U} = \{\bar{C}\}$ serves as an analytical building block). We establish three properties of $F$ that make \eqref{eq:r-seed} tractable: a closed-form optimal perturbation (\cref{lemma:optimal_perturbation}), NP-hardness of the selection problem, and monotone weak submodularity (\cref{lemma:submodularity}).

\begin{lemma}
\label{lemma:optimal_perturbation}
For a fixed set $S \subseteq V$, the optimal perturbation is 
$\delta^*_S = -Z^{-1}_{SS} Z_{S:} s$, and can be computed in polynomial time. The resulting benefit function is $F(S, L, C) = s^T P_S s$, where $P_S = Z_{:S} Z^{-1}_{SS} Z_{S:}$ and $Z = M \odot C$ with $M = (I+L)^{-2}$.
\end{lemma}

\begin{lemma}
\label{lemma:submodularity}
The opinion seeding problem $\max_{S:\ |S| \le b} F(S, L, C)$ is NP-hard. Moreover, $F(S, L, C)$ is monotone and weakly submodular with submodular ratio $\gamma = 1 / \kappa(Z)$ where $\kappa(Z)$ is the condition number of $Z$. Consequently, the greedy hill-climbing algorithm achieves a $(1 - e^{-\gamma})$-approximation.
\end{lemma}

For each fixed $C$, this weak submodularity lets the greedy algorithm approximate the inner maximization over $S$ to within $(1 - e^{-\gamma})$ \citep{das2011submodular,nemhauser1978analysis}, serving as the oracle inside an active set algorithm that adapts the robust Saturate framework of \citet{krause2008robust} for a bicriteria guarantee.

\xpar{Seeding algorithm} The algorithm maintains an active set $\mathcal{A}_k$ of adversarial correlation matrices encountered so far, initialized as $\mathcal{A}_0 = \{\bar{C}\}$. At each outer iteration $k = 1, \ldots, K$: 

\medskip

\begin{itemize}
    \item \emph{Step 1: Seeding oracle.} First, we apply the Saturate algorithm of \citet{krause2008robust} to the collection of benefit functions $\{F_{C}(\cdot) : C \in \mathcal{A}_{k-1}\}$ with seeding budget $b_k = \left ( 1 + \frac {\log (k / \varepsilon)}{\gamma'_k} \right ) b$ to obtain a seed set $\hat S$ that is robust to all scenarios in the current active set, that is, obtain $\hat S_k = \mathrm{Saturate} \left(\{F_C : C \in \mathcal{A}_{k-1}\}\right).$ The weak submodularity ratio is $\gamma_k' = \min_{C \in \mathcal A_k} \frac {1} {\kappa(M \odot C)}$.
    \item \emph{Step 2: group structure update.}  Then we solve the inner maximization to find the most violated scenario given $\hat S_k$, that is $\hat{C}_k = \arg\min_{C \in \mathcal{U}} F({\hat S}_k, L, C),$ via projected gradient descent over $\mathcal{U}$. If $F(\hat S_k, L, \hat{C}_k) < \min_{C \in \mathcal{A}_{k-1}} F(\hat S_k, L, C)$, set $\mathcal{A}_k = \mathcal{A}_{k-1} \cup \{\hat{C}_k\}$; 
    otherwise set $\mathcal{A}_k = \mathcal{A}_{k-1}$.
\end{itemize}

\medskip

Finally, we output $\hat S_K$ corresponding to the final active set  $\mathcal{A}_K$, and the corresponding correlation matrix $\hat C_K$. In the sequel, we prove the approximation ratio of our algorithm by adapting the Saturate algorithm from \citet{krause2008robust}: 

\begin{lemma}
\label{lem:saturate-seeding}

Let $\hat S = \hat S_K$ be the resulting set after $K$ outer iterations, targeting a robust benefit value $\alpha \le \opt_b = \max_{|S| \le b} \min_{C \in \mathcal A_K} F(S, L, C)$ within a tolerance $\varepsilon > 0$. The output seed set satisfies $\min_{C \in \mathcal A_K} F(\hat{S}, L, C) \ge \alpha - \varepsilon$, with size bounded by $|\hat{S}| \le \left( 1 + \frac{\log (K \alpha / \varepsilon)}{\gamma_K'} \right) b$, where $\gamma_K' = \min_{C \in \mathcal A_K} \frac{1}{\kappa(M \odot C)}$ is the worst-case submodularity ratio.

\end{lemma}

Then, we provide a guarantee for the final algorithm:

\begin{theorem} \label{theorem:saturate-final}
    Suppose the outer active set algorithm terminates at iteration $K$ when the inner maximization oracle certifies that no matrix $C \in \mathcal{U}$ violates the active set's worst-case evaluation by more than an outer tolerance $\varepsilon_C > 0$. Specifically, the algorithm halts when: $$F(\hat{S}_K, L, \hat{C}_K) \ge \min_{C \in \mathcal{A}_{K-1}} F(\hat{S}_K, L, C) - \varepsilon_C,$$ where $\hat{C}_K = \arg\min_{C \in \mathcal{U}} F(\hat{S}_K, L, C)$.Then, the final seed set $\hat{S}_K$ achieves the following global robust guarantee against the entire uncertainty set:$$\min_{C \in \mathcal{U}} F(\hat{S}_K, L, C) \ge \alpha - \varepsilon - \varepsilon_C.$$
\end{theorem}

\xpar{Regularized Seeding} The closed-form perturbation $\delta_S^* = -Z_{SS}^{-1}Z_{S\bar{S}}s_{\bar{S}}$ 
in \cref{lemma:optimal_perturbation} is unconstrained: it imposes no bound on the magnitude of the recommended opinion shift, and the perturbed opinions $s_S + \delta_S^*$ need not lie in any operationally feasible domain. In practice, platforms operate under intervention budgets, content boosts are bounded in intensity, and individual users cannot be moved arbitrarily, that is a constraint $\| \delta_{S} \| \le 2 / \nu$ is imposed for some reasonable budget $\nu > 0$. We show here that adding a quadratic penalty on the perturbation magnitude yields a tractable constrained variant that preserves the Schur complement structure of the benefit function and therefore retains monotone weak submodularity. Specifically, for a fixed seed set $S \subseteq V$ and budget parameter $1 / \nu > 0$, consider the Tikhonov regularized version:
\begin{equation}
\min_{\delta: \mathrm{supp}(\delta) \subseteq S}  
g(s + \delta,  L,  C) + \nu \|\delta \|^2.
\label{eq:reg-seed}
\end{equation}

\noindent The parameter $\nu$ penalizes large opinion shifts, trading off disparity reduction against intervention effort. Setting $\nu \to 0$ recovers the unconstrained problem of \cref{lemma:optimal_perturbation}.

\begin{lemma}
\label{lemma:reg-seed}
For fixed $S$, $L$, $C$, and $\nu \ge 0$, the unique minimizer of \eqref{eq:reg-seed} constrained to $S$ is $\delta_S^*(\nu) = -(Z_{SS} + \nu I)^{-1} Z_{S:} s$, which satisfies $\|\delta_S^*(\nu)\| \le 2/\nu$ for all $\nu > 0$. The resulting regularized benefit function is $F_\nu(S, L, C) = g(s, L, C) - g(s + \delta_S^*(\nu), L, C) = s^T P_S^\nu s$, where $P_S^\nu = Z_{:S}(Z_{SS} + \nu I)^{-1}Z_{S:}$. Moreover, $F_\nu(\cdot, L, C)$ is monotone and weakly submodular in $S$ for every $\nu \ge 0$ with $\gamma_\nu = 1/\kappa(Z + \nu I)$, and greedy hill-climbing achieves a $(1 - e^{-\gamma_\nu})$-approximation to $\max_{S:|S|\le k} F_\nu(S, L, C)$.
\end{lemma}

\subsection{Experiments} \label{app:seeding_experiments}

We conduct opinion seeding experiments on the same three real-world networks used in \cref{sec:experiments}: Polblogs, Reddit, and Twitter. For each network, the platform 
constructs a nominal group structure $\bar{r} = \mathrm{sign}(v_2)$ via spectral clustering, and the uncertainty set $\mathcal{U}(\rho)$ is parameterized by $\rho = 0.2$ throughout. We set the nominal seeding budget to $b = 100$ and the regularization parameter to $\nu = 2$, allowing perturbations of magnitude at most $1/2$. We report the benefit function $F(S, L, C)$, and we compare the greedy hill-climbing  algorithm against three uninformed baselines: random selection, max-degree selection, and PageRank-based selection.

\xpar{Nominal seeding oracle and baselines} \cref{fig:opinion_seeding_oracles} reports the first iteration of the robust algorithm (with $\mathcal{A}_0 = \{\bar{C}\}$), i.e.\ the nominal oracle operating against $\bar C$. The benefit increases monotonically over greedy steps (consistent with the weak submodularity of \cref{lemma:submodularity}), with runtimes of tens of seconds for networks with thousands of users, and the normalized interventions $\delta_S/\|\delta_S\|$ are heterogeneous across $\ge 90\%$ of seeded users, with no systematic suppression or amplification, as for recommendation reweighing (\cref{fig:impact}). The greedy algorithm consistently beats all three baselines (random, max-degree, PageRank), often substantially and most so on Polblogs, where the community structure is clearest; max-degree and PageRank track each other but stay well below greedy, confirming that selecting high-influence users without regard for disparity reduction is insufficient.

\begin{figure}[t]
    \centering
    \includegraphics[width=0.8\linewidth]{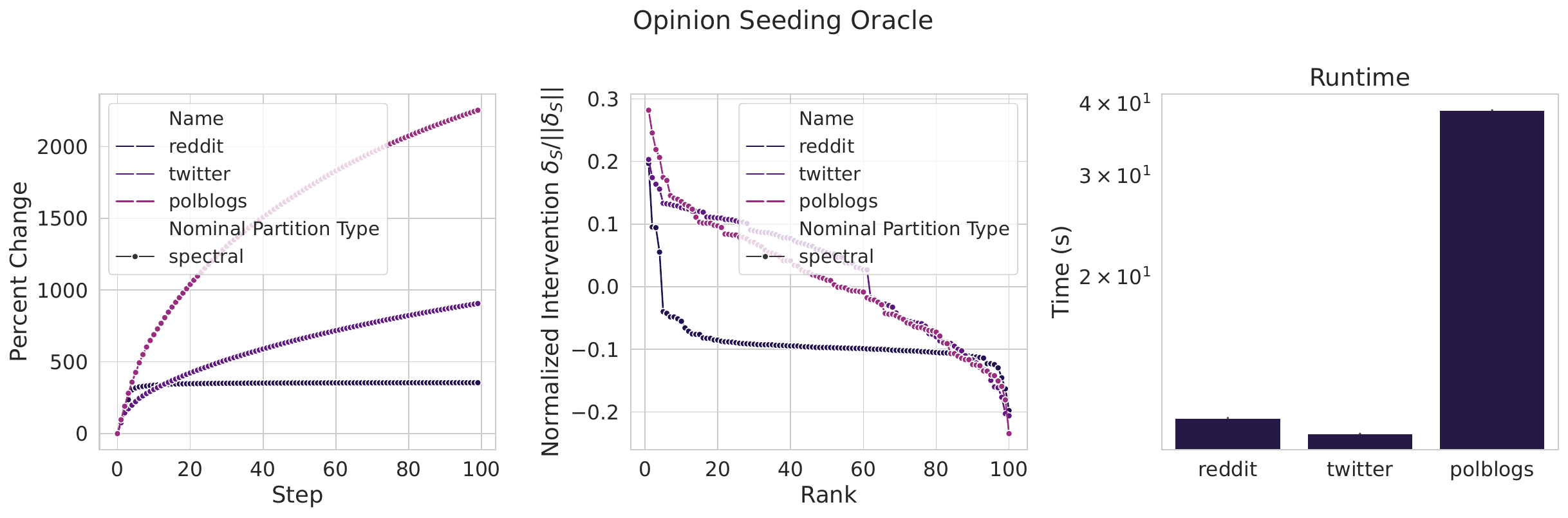}
    \includegraphics[width=0.8\linewidth]{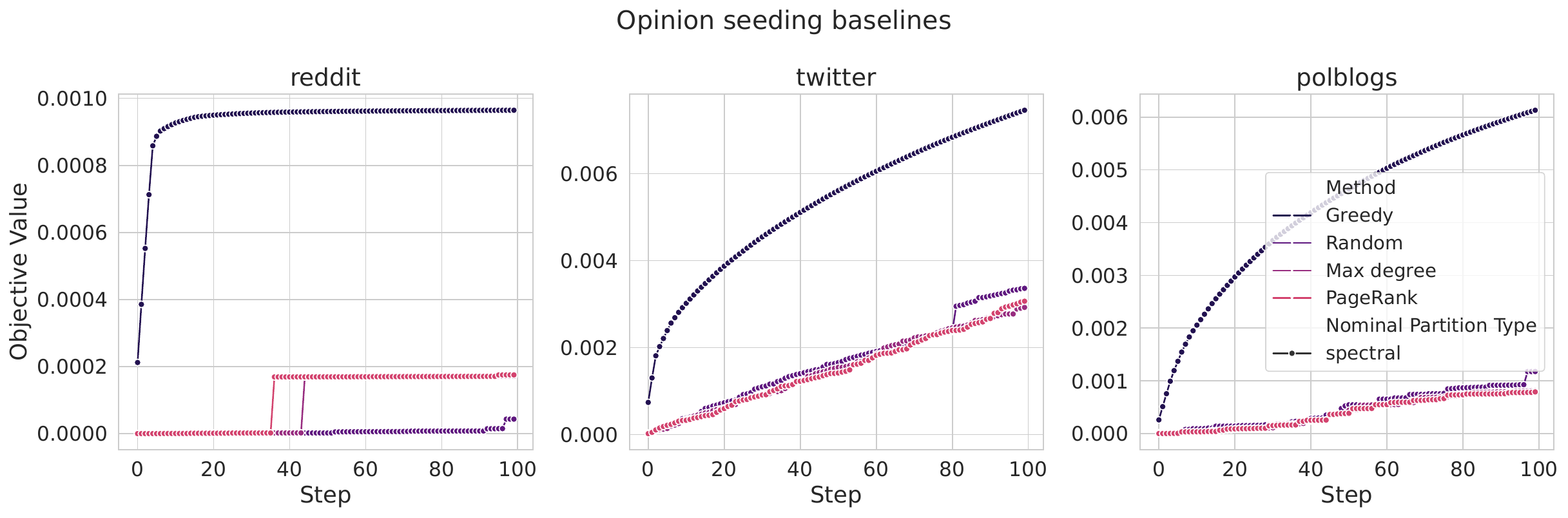}
    \caption{Top: Opinion seeding oracle output on the first step of the robust algorithm, i.e., with $\mathcal A_0 = \{ \bar C \}$). We report the percent change in the benefit objective for each of the three datasets, as well as the normalized opinions $\delta_S / \| \delta_S \|$ at the end. The algorithm runs in the order of tens of seconds for networks with thousands of users. Bottom: Comparison of the greedy algorithm with baselines (random, max degree, PageRank) on the first step of the robust algorithm. We observe that the greedy algorithm outperforms all baselines, often by a large margin.}
    \label{fig:opinion_seeding_oracles}
\end{figure}

\xpar{Robust seeding algorithm} \cref{fig:robust_seeding} reports the final output after at most $K=3$ outer iterations ($b=100$, $\rho=0.2$). The benefit converges within one iteration as the active set grows to cover the most adversarial group structures, and the final intervention $\delta_{\hat S}/\|\delta_{\hat S}\|$ is large for fewer than $10\%$ of seeded users. This sparsity is operationally desirable: the platform applies strong shifts to only a few targeted accounts while the rest receive modest nudges, reducing the intervention's surface area and improving auditability.

\begin{figure}[t]
    \centering
    \includegraphics[width=0.8\linewidth]{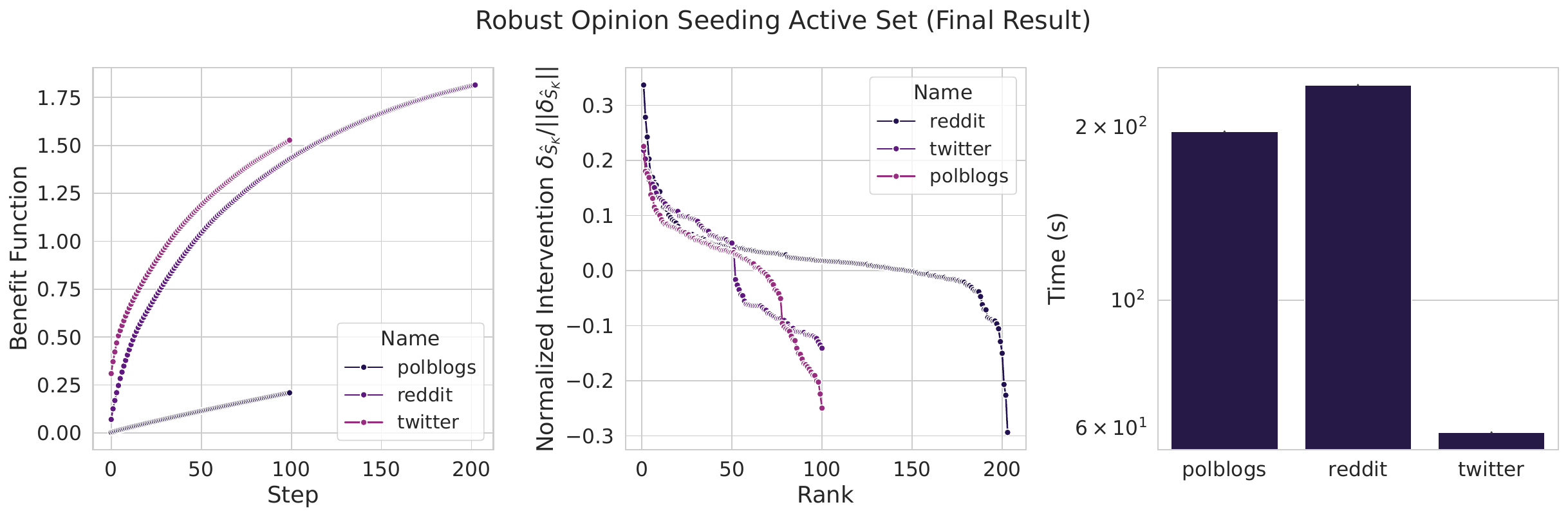}
    \caption{Last iteration of the robust opinion seeding oracle for a per-oracle seeding budget of $b = 100$, error $\rho = 0.2$, $K = 3$ outer iterations, and regularizer $\nu = 2$. We report the benefit objective for each of the three datasets, as well as the normalized opinions $\delta_{\hat S} / \| \delta_{\hat S} \|$ at the end of the robust opinion seeding algorithm. The algorithm runs in the order of $10^2-10^3$ for all the datasets. We observe that the intervention $\delta_{\hat S}$ is large for less than 10\% of the seeded users.}
    \label{fig:robust_seeding}
\end{figure}

\section{Proofs}\label{app:proofs}

\subsection{Proof of \cref{theorem:gap}}
Let $\{v_k,\lambda_k\}_{k=1}^n$ be the eigenbasis of $L$ and $\hat S_k = s^T v_k$. Since $s = \bar r/\sqrt n$ and the partition is balanced, $s \perp v_1$, so $\hat S_1 = 0$ and $\sum_{k\ge2}\hat S_k^2 = 1$. Hence, the polarization satisfies
\begin{equation}
    g(s, L, \one\one^T) = \sum_{k=2}^n \frac{\hat S_k^2}{(1+\lambda_k)^2} \le \frac{1}{(1+\lambda_2)^2},
    \label{eq:pol-bounds}
\end{equation}
using $\lambda_k \ge \lambda_2$ for $k \ge 2$. For the disparity, in the flipping model $r = \bar r \odot \epsilon$ with i.i.d.\ signs $\epsilon_i$ and $\Pr(\epsilon_i = +1) = 1-\rho$; since $s_i\bar r_i = 1/\sqrt n$, we have $s \odot r = \epsilon/\sqrt n$, so $$g(s,L,C(\rho)) = \tfrac1n\,\mathbb E_\epsilon[\epsilon^T M \epsilon] = \tfrac1n\Big[\tr(M) + (1-2\rho)^2\textstyle\sum_{i\neq j}M_{ij}\Big],$$
using $\mathbb E[\epsilon_i^2]=1$ and $\mathbb E[\epsilon_i\epsilon_j] = (1-2\rho)^2$ for $i \neq j$. Because $v_1 = \one/\sqrt n$ with $\lambda_1 = 0$, $\one^T M\one/n = 1$, i.e.\ $\tr(M) + \sum_{i\neq j}M_{ij} = n$; substituting and using $1 - 4\rho(1-\rho) = (1-2\rho)^2$ gives $g(s,L,C(\rho)) = 1 - \tfrac{4\rho(1-\rho)}{n}\sum_{i\neq j}M_{ij} \ge (1-2\rho)^2$, where the last step uses $\sum_{i\neq j}M_{ij} \le n$ (since $\tr(M)\ge0$). Dividing this lower bound by the polarization upper bound \eqref{eq:pol-bounds} yields $g(s,L,C(\rho))/g(s,L,\one\one^T) \ge (1-2\rho)^2(1+\lambda_2)^2$, as claimed.

\subsection{Proof of \cref{prop:surrogate_upper_bound}}

\xpar{Upper bound} Since $L \succeq 0$, the eigenvalues of $X = (I+L)^{-1}$ lie in $(0,1]$, so $X \preceq I$ and $X - X^2 = X(I-X) \succeq 0$. By the Schur product theorem, $(X-X^2)\odot C \succeq 0$ for any $C \succeq 0$, i.e.\ $s^T(X^2\odot C)s \le s^T(X\odot C)s$ for all $s$, which is $g(s,L,C) \le \tilde g(s,L,C)$.

\xpar{Convexity} The map $X(L) = (I+L)^{-1}$ is matrix convex on $\mathcal L$ \citep{musco2018minimizing}: for $\lambda\in[0,1]$, $\lambda(I+L_1)^{-1} + (1-\lambda)(I+L_2)^{-1} \succeq (I+\lambda L_1 + (1-\lambda)L_2)^{-1}$. Taking the Hadamard product with $C \succeq 0$ preserves the PSD order (Schur), and the induced quadratic form in $s$ shows that $\tilde g(s,L,C) = s^T(X(L)\odot C)s$ is convex in $L$.

\xpar{Spectral error bound} Let $\epsilon(s,L,C) = \tilde g(s,L,C) - g(s,L,C)$ and $D = X - X^2 \succeq 0$, whose eigenvalues are $f(\lambda_i) = \lambda_i/(1+\lambda_i)^2$, so $D \preceq \mu_{\max} I$ with $\mu_{\max} = \max_i f(\lambda_i)$. Since $(\mu_{\max}I - D)\odot C \succeq 0$ and $I\odot C = I$, we get $D\odot C \preceq \mu_{\max}I$, hence $0 \le \epsilon(s,L,C) = s^T(D\odot C)s \le \mu_{\max}$ for $\|s\|=1$. Finally $f'(\lambda) = (1-\lambda)/(1+\lambda)^3$ vanishes only at $\lambda = 1$ (a maximum), so $f(\lambda) \le f(1) = 1/4$ and $\mu_{\max} \le 1/4$.

\xpar{Tightness} The bound is tight: for the path on three nodes with $s = (1,0,-1)^T/\sqrt2$, whose Laplacian has eigenvalues $\{0,1,3\}$, the polarization setting $\bar C = \one\one^T$ gives $g(s,L,\one\one^T) = \tfrac{1}{(1+1)^2}\|s\|^2 = \tfrac14$ and $\tilde g(s,L,\one\one^T) = \tfrac{1}{1+1}\|s\|^2 = \tfrac12$, so $\tilde g - g = \tfrac14 = \mu_{\max}$, attained at $\lambda_2 = 1$.

\subsection{Proof of \cref{theorem:fj_model_link_update}}

\xpar{Leverage score for the disparity $g$} Set $X = (I+L)^{-1}$ and let $b_e$ be the incidence vector of edge $e$. Differentiating the resolvent (Sherman--Morrison) gives $\partial X^2/\partial w_e = -X(Xb_eb_e^T + b_eb_e^T X)X = -Q_e$, following \citet{wang2024relationship}. Hence for the conditional disparity $f = y^T X^2 y$ with $y = s\odot r$, we have $\partial f/\partial w_e = -y^T Q_e y$, and by linearity of expectation $\partial g/\partial w_e = \mathbb E_r[-y^T Q_e y] = -s^T(Q_e\odot C)s$.

\xpar{Surrogate leverage} Likewise $\partial X/\partial w_e = -(Xb_e)(Xb_e)^T = -\tilde Q_e$, so $\partial \tilde f/\partial w_e = -y^T\tilde Q_e y$ and $\partial\tilde g/\partial w_e = -s^T(\tilde Q_e\odot C)s$.

\xpar{Robust leverage} For the robust surrogate $\phi(L) = \max_{C \in \mathcal U(\rho)}\lambda_{\max}(X\odot C)$, Danskin's theorem gives the gradient at the inner maximizer $C^*$: with $v$ the leading eigenvector of $X\odot C^*$,
\begin{equation*}
\frac{\partial\phi}{\partial w_e} = -v^T(\tilde Q_e\odot C^*)v =: -\tilde\beta_e^{\mathrm{rob}}, \qquad \tilde Q_e = Xb_eb_e^T X.
\end{equation*}

\subsection{Proof of \cref{theorem:link_prediction_approximation}}

This appendix establishes the approximation and runtime guarantees of the robust recommendation reweighing algorithm of \cref{sec:link_recommendation}. We recall, that, $\mathcal{L}$ denotes the convex polytope of admissible Laplacians,
and $\mathcal{U}(\rho)$ denotes the uncertainty set. We write
$X(L)=(I+L)^{-1}$, $M(L)=(I+L)^{-2}$, and for any normalized
$s\in\mathbb{R}^n$, $C\in\mathcal{U}(\rho)$, define the true and
surrogate disparities by
\begin{equation*}
   g(s,L,C) = s^T(M(L)\odot C)s, \qquad
   \tilde g(s,L,C) = s^T(X(L)\odot C)s.
\end{equation*}
The robust true and surrogate objectives are
\begin{equation*}
   \Psi(L)  =  \max_{C\in\mathcal{U}} h(L, C),
   \qquad
   \phi(L)  =  \max_{C\in\mathcal{U}} \tilde h(L, C), \qquad \tilde h(L, C) = \lambda_{\max} (X(L) \odot C)
\end{equation*}
so that $\opt = \min_{L\in\mathcal{L}}\Psi(L)$ is the optimum of
the original robust problem (R-Link) and $\opt_\phi =
\min_{L\in\mathcal{L}}\phi(L)$ is the optimum of the surrogate problem. 

 
 

\subsubsection{Helper Lemmas}\label{app:structural}
 
\begin{lemma}\label{lem:phi}
The function $\phi:\mathcal{L}\to\mathbb{R}$ is convex,
$0 \le \phi(L) \le 1$ for every $L\in\mathcal{L}$, and is
$G_\phi$-Lipschitz in Frobenius norm with $G_\phi \le 1$.
\end{lemma}
 
\xpar{Proof} For fixed $s$ with $\|s\|=1$ and fixed $C\in\mathcal{U}$, the map $L\mapsto \tilde g(s,L,C)$ is matrix-convex by Proposition~1 (combining matrix-convexity of $L\mapsto(I+L)^{-1}$ with the Schur product theorem applied to $C\succeq 0$). The pointwise maximum of a family of convex functions is convex, so $\phi$ is convex. For the boundedness, note that $L\succeq 0$ implies all eigenvalues of $X(L)$ lie in $(0,1]$, so $X(L) \preceq I$. The Schur product theorem yields $X(L)\odot C \preceq I \odot C = I$, hence $\lambda_{\max}(X(L)\odot C) \le 1$, and non-negativity follows from $X(L)\odot C \succeq 0$. For the Lipschitz bound, Danskin's theorem combined with Proposition~2
gives, for each $e\in E$,
\begin{equation*}
   \frac{\partial \phi}{\partial w_e}(L)
   = -v^T(\tilde Q_e \odot C^*)v,
   \qquad
   \tilde Q_e = X(L) b_e b_e^T X(L),
\end{equation*}
where $(C^*,v)$ are the inner maximizers. By Danskin's theorem and the
chain rule, the subgradient of $\phi$
with respect to $L$ is $\nabla_L\phi = -X(L)\,(C^*\odot vv^T)\,X(L)$.
Hence $\|\nabla_L\phi\|_F \le \|X(L)\|^2\,\|C^*\odot vv^T\|_F \le 1$,
since $\|X(L)\|\le 1$ and $\|C^*\odot vv^T\|_F^2 = \sum_{i,j}(C^*_{ij})^2 v_i^2 v_j^2 \le \bigl(\textstyle\sum_i v_i^2\bigr)^2 = 1$, using $|C^*_{ij}|\le 1$ and $\|v\|=1$. Therefore $\phi$ is $G_\phi$-Lipschitz in Frobenius norm with $G_\phi\le 1$.
 
\begin{lemma}\label{lem:gap}
For every $L\in\mathcal{L}$,
$
   0 \le \phi(L) - \Psi(L) \le \mu_{\max}(L) \le 1/4,
$
where $\mu_{\max}(L) = \max_i \lambda_i(L)/(1+\lambda_i(L))^2$.
\end{lemma}
 
\xpar{Proof} By Proposition~1, $\tilde g(s,L,C) - g(s,L,C) \le \mu_{\max}(L)$ for every $(s,C)$ with $\|s\|=1$. Hence
\begin{equation*}
   \phi(L) - \Psi(L)
   = \max_{s,C}\tilde g(s,L,C) - \max_{s,C} g(s,L,C)
   \le \max_{s,C}\bigl[\tilde g(s,L,C)-g(s,L,C)\bigr]
   \le \mu_{\max}(L).
\end{equation*}
The bound $\mu_{\max}(L)\le 1/4$ is the global maximum of
the function $f(\lambda) = \lambda/(1+\lambda)^2$ on $[0,\infty)$, attained at $\lambda=1$.
 
\begin{lemma}\label{lem:bracket}
$\opt \le \opt_\phi \le \opt + \mu_{\max}$.
\end{lemma}
 
\xpar{Proof} By Lemma~\ref{lem:gap}, $\phi \ge \Psi$ pointwise, so
$\opt_\phi \ge \opt$. Conversely, letting
$L^* = \arg\min_L \Psi(L)$,
$\opt_\phi \le \phi(L^*) \le \Psi(L^*) + \mu_{\max} =
\opt + \mu_{\max}$.
 
\begin{lemma}\label{lem:diameter}
$D = \operatorname{diam}_F(\mathcal{L}) \le 2 \bar{w}\sqrt{|E|}$.
\end{lemma}
 
\xpar{Proof} For $L_1,L_2\in\mathcal{L}$ with edge weights $w^{(1)},w^{(2)}$, by the triangle inequality we get: $$D^2 = \|L_1-L_2\|_F^2 \le \sum_e |w_e^{(1)}-w_e^{(2)}|^2\|b_eb_e^T\|_F^2 \le 4\sum_e |w_e^{(1)}-w_e^{(2)}|^2 \le 4|E|\bar{w}^2.$$
 
\subsubsection{Inner Loop Convergence}\label{app:inner} 
 
Fix an outer iteration $k$ and let $C_k^*\in\mathcal{A}_k$ denote the
worst-case correlation matrix selected by Step~2 at the beginning of
that outer iteration. Define $h_k(L)  =  \lambda_{\max}\bigl(X(L)\odot C_k^*\bigr).$
By Lemma~\ref{lem:phi} applied with a single $C$ in place of the
maximum, $h_k$ is convex on $\mathcal{L}$ and $G_\phi$-Lipschitz. The
pair-swap update at inner step $t$ is
\begin{equation*}
   L_{k,t+1}
    =  L_{k,t}
    +  \eta_t' \frac {m} {b} \bigl(b_{e_t^+}b_{e_t^+}^T - b_{e_t^-}b_{e_t^-}^T\bigr),
\end{equation*}
for $|B_t| = b$ with step size $\eta_t' = \min\{\eta_t,
\bar w - w_{e_t^+}, w_{e_t^-}\}$, ensuring $L_{k,t+1}\in\mathcal{L}$. The
trace is preserved because $\operatorname{tr}(b_eb_e^T)=2$ for every
$e$.

\begin{lemma} \label{lem:inner-rate}
Let $L_k^* = \arg\min_{L\in\mathcal{L}} h_k(L)$, and let
$\bar L_{k,T} = \frac{1}{T}\sum_{t=1}^T L_{k,t}$. With step sizes
$\eta_t = D/(G_\phi\sqrt{t})$,
\begin{equation}\label{eq:inner-rate}
   \mathbb{E}\bigl[h_k(\bar L_{k,T})\bigr] - h_k(L_k^*)
    \le  O \left ( \frac{D G_\phi}{\sqrt{T}} \right ).
\end{equation}

\end{lemma}

\xpar{Proof} The objective $h_k(L) = \lambda_{\max}(X(L) \odot C_k^*)$ is convex and $G_\phi$-Lipschitz (\cref{lem:phi}), and its subgradient $\xi_t \in \partial h_k(L_{k,t})$ has edge components equal to the negative robust leverage scores, $(\xi_t)_e = -\tilde{\beta}_e^{\mathrm{rob}}$. Because the trace constraint $\sum_e w_e = 2m$ makes any feasible displacement $L_{k,t} - L_k^*$ a convex combination of elementary pair-swaps, the best pair-swap over $E$ captures an $\Omega(1/|E|)$ fraction of the optimality gap; sampling a random batch $B_t$ and taking pair $d_t = \tfrac{m}{b}(b_{e_t^+}b_{e_t^+}^T - b_{e_t^-}b_{e_t^-}^T)$ preserves this in expectation, so $\mathbb{E}[\langle \xi_t, -d_t \rangle \mid L_{k,t}] \ge C\,(h_k(L_{k,t}) - h_k(L_k^*))$ for a constant $C>0$. Treating $d_t$ as a stochastic pseudo-gradient with $\mathbb{E}\|d_t\|_F^2 \le G_\phi^2$, the standard projected-subgradient distance recurrence with step size $\eta_t = D/(G_\phi\sqrt{t})$, summed over $t$ and combined with Jensen's inequality for the averaged iterate $\overline L_{k,T}$, yields $\mathbb{E}[h_k(\overline{L}_{k,T})] - h_k(L_k^*) \le O\!\left( D G_\phi/\sqrt{T} \right)$.

\subsubsection{Outer Loop Convergence}\label{app:outer}
 
Define the partial worst case at iteration $k$, $\hat\phi_k(L)  =  \max_{C\in\mathcal{A}_k} \tilde h(L, C),$ such that $\hat\phi_k(L) \le \phi(L)$ for all $L \in \mathcal L$.

\begin{lemma}\label{lem:outer-rate}

Suppose the algorithm reaches saturation at outer iteration $K$: the separation oracle (Step 2) certifies that no $C \in \mathcal{U}$ improves on the active set maximum by more than $\varepsilon_C$, meaning: $$\hat \phi_K(L) \le \phi(L) \le \hat \phi_K(L) + \varepsilon_C,$$ where $\overline{L}_{K,T}$ is the averaged iterate returned by the inner loop. Then the expected gap to the true robust surrogate optimum $\opt_\phi$ is bounded by: $$\mathbb{E}[\phi(\overline{L}_{K,T})] - \opt_\phi \le O \left( \frac{D G_\phi}{\sqrt{T}} \right) + \varepsilon_C$$

\end{lemma}

\xpar{Proof} The saturation certificate at $\overline{L}_{K,T}$ gives, deterministically, $\phi(\overline{L}_{K,T}) \le \hat{\phi}_K(\overline{L}_{K,T}) + \varepsilon_C$, so taking expectations over the inner loop's batches, $\mathbb{E}[\phi(\overline{L}_{K,T})] \le \mathbb{E}[\hat{\phi}_K(\overline{L}_{K,T})] + \varepsilon_C$. Since the inner loop runs batch-greedy subgradient descent on $\hat{\phi}_K$ for $T$ steps, \cref{lem:inner-rate} gives $\mathbb{E}[\hat{\phi}_K(\overline{L}_{K,T})] \le \min_{L \in \mathcal{L}} \hat{\phi}_K(L) + O(D G_\phi/\sqrt{T})$. Finally, $\mathcal{A}_K \subseteq \mathcal{U}$ implies $\hat{\phi}_K \le \phi$ pointwise, so $\min_{L \in \mathcal{L}} \hat{\phi}_K(L) \le \min_{L \in \mathcal{L}} \phi(L) = \opt_\phi$. Combining these inequalities yields $\mathbb{E}[\phi(\overline{L}_{K,T})] - \opt_\phi \le O(D G_\phi/\sqrt{T}) + \varepsilon_C$.

\subsubsection{Proof of the Main Result}\label{app:assembly}
 
To prove the main result, we decompose the expected optimality gap as $\ev{}{\Psi(\overline L_{K,T})} - \opt = [\ev{}{\Psi(\overline L_{K,T})} - \ev{}{\phi(\overline L_{K,T})}] + [\ev{}{\phi(\overline L_{K,T})} - \opt_\phi] + [\opt_\phi - \opt]$. The first term is $\le 0$ since $\Psi \le \phi$ pointwise (\cref{lem:gap}), the second is $\le O(D G_\phi/\sqrt{T}) + \varepsilon_C$ by \cref{lem:outer-rate}, and the third is $\le \mu_{\max}$ by \cref{lem:bracket}. Substituting $D \le 2\bar w\sqrt{|E|}$ (\cref{lem:diameter}) and $G_\phi \le 1$ (\cref{lem:phi}) gives the claimed bound $\ev{}{\Psi(\overline L_{K,T})} - \opt \le O(\bar w\sqrt{|E|}/\sqrt{T}) + \varepsilon_C + \mu_{\max}$.

\subsection{Proofs of \cref{lemma:optimal_perturbation,lemma:submodularity}}

\xpar{Definition of the benefit function $F(S, L, C)$} For a seed set $S$, the benefit function is the disparity reduction from optimally seeding $S$: $$F(S, L, C) = g(s, L, C) - \min_{\delta : \supp (\delta) \subseteq S} g(s + \delta, L, C).$$

We drop the dependence on $L,C$, writing $F(S) = F(S,L,C)$, and partition $Z = M\odot C$, $\delta$, and $s$ into $S$ and $\bar S$ blocks with $\delta_{\bar S}=0$. Since $Z\succeq0$, the objective $$g(s+\delta) = s^T Z s + 2\delta_S^T\big(Z_{SS}s_S + Z_{S\bar S}s_{\bar S}\big) + \delta_S^T Z_{SS}\delta_S = s^T Z s + 2\delta_S^T Z_{S:}s + \delta_S^T Z_{SS}\delta_S$$ is convex in $\delta_S$, so its first-order condition $Z_{SS}\delta_S + Z_{S:}s = 0$ gives the unique minimizer $\delta_S^* = - Z_{SS}^{-1} Z_{S:} s$ and  $\delta_{\bar S} = 0$.
Substituting back yields $F(S) = s^T P_S s$ with $P_S = Z_{:S} Z_{SS}^{-1} Z_{S:}$, so the seeding problem is equivalent to solving $\max_{S : |S| \le b} F(S)$.

\xpar{NP-hardness} We reduce from Maximum $k$-Coverage: given a universe $U = \{e_1, \dots, e_m\}$ and subsets $\mathcal{V} = \{V_1, \dots, V_n\}$ with $V_i \subseteq U$, select $b$ subsets maximizing the number of covered elements. Build a bipartite network with a candidate node per subset $V_i$ (set $V_{\mathrm{cand}}$), a target node per element $e_j$ (set $V_{\mathrm{targ}}$), and a sink $v_0$ for connectivity and the zero-mean constraint. Connect $V_i$ to $e_j$ with large weight $W$ iff $e_j \in V_i$, and every node to $v_0$ with weight $\epsilon > 0$. Set $C = \one\one^T$ (so $Z = M = (I+L)^{-2}$) and opinions $s_j = 1$ on targets, $s_i = 0$ on candidates, $s_0 = -m$ on the sink (ensuring $\one^T s = 0$, $\|s\|^2 = 1$), and restrict seeds to $S \subseteq V_{\mathrm{cand}}$. The benefit $F(S) = s^T (Z_{:S} Z_{SS}^{-1} Z_{S:}) s$ is the reduction in squared equilibrium opinions from optimally intervening on $S$. As $W \to \infty$ connected nodes share an equilibrium opinion, and since candidates start at $0$, seeding $V_i$ neutralizes the opinions of all targets adjacent to it; thus as $W \to \infty$, $\epsilon \to 0$, $F(S)$ counts the unique targets covered by $S$. Maximizing $F(S)$ over $|S| \le b$ therefore solves Maximum $k$-Coverage, which is strongly NP-hard, so opinion seeding is NP-hard.

\xpar{Monotone Weak Submodularity of $F(S)$} Recall $F(S) = s^T P_S s$ with $P_S = Z_{:S} Z_{SS}^{-1} Z_{S:}$. Monotonicity is immediate: if $A \subseteq B$ then $P_B \succeq P_A$, so $F(B) \ge F(A)$. For weak submodularity\footnote{Generally, one may be tempted to think that $F(S, L, C)$ is a submodular function; this is not correct, as one can construct instances where the submodularity property fails.}, define the conditioned matrix $Z^{|T|} = Z - Z_{:,T} Z_{TT}^{-1} Z_{T,:}$ for $T \subseteq [n]$. By \cref{lemma:rankone-marginal}, with $v_u = s^T Z_{:,u}^{|T|}$ the joint and summed singleton gains are $$F(T \cup S) - F(T) = v^T \big( Z_{SS}^{|T|} \big)^{-1} v, \qquad \sum_{u \in S}\big(F(T \cup \{u\}) - F(T)\big) = v^T \big[ \diag( Z_{SS}^{|T|} ) \big]^{-1} v.$$

The submodularity ratio is therefore the Rayleigh quotient, which under the substitution $w = A^{-1/2}v$ with $A = Z_{SS}^{|T|}$, $D = \diag(A)$ becomes $$\gamma = \min_{v \neq \zero} \frac{v^T [ \diag( Z_{SS}^{|T|} ) ]^{-1} v}{v^T ( Z_{SS}^{|T|} )^{-1} v} = \lambda_{\min}\big( D^{-1/2} A D^{-1/2} \big) \ge \frac{\lambda_{\min}(A)}{\max_i D_{ii}}.$$
By the eigenvalue interlacing theorem $\lambda_{\min}(A) \ge \lambda_{\min}(Z)$, and since conditioning reduces variance $\max_i D_{ii} = \max_{u} Z_{uu}^{|T|} \le \max_{u} Z_{uu} \le \lambda_{\max}(Z)$. Hence $\gamma \ge \lambda_{\min}(Z)/\lambda_{\max}(Z) = 1/\kappa(Z)$. As $F$ is monotone and weakly submodular with ratio $\gamma$, greedy hill-climbing yields a $(1 - e^{-\gamma})$-approximation due to \citet{das2011submodular}.

\subsection{Proof of \cref{lem:saturate-seeding}}

The platform solves $\max_{|S| \le b} \min_{C \in \mathcal{U}} F(S, L, C)$. For each $C$, $F(\cdot,L,C)$ is monotone and weakly submodular with ratio $\gamma_C = 1/\kappa(M\odot C)$, so we extend the Saturate algorithm of \citet{krause2008robust} from submodular to weakly submodular functions. Saturate does binary search for a target $\alpha \in [0,\opt_b]$ by solving the min-cover problem $\min |S|$ subject to $F_C^\alpha(S) \ge \alpha$ for all $C \in \mathcal A_K$, where $F_C^\alpha(S) = \min(F_C(S),\alpha)$ is the truncated benefit and $|\mathcal A_K| \le K$. Equivalently, with the averaged function $\hat F^\alpha(S) = \sum_{C \in \mathcal A_K} F_C^\alpha(S)$, the target holds iff $\hat F^\alpha(S) \ge |\mathcal A_K|\alpha$.

Truncation preserves weak submodularity (each $F_C^\alpha$ has ratio at least $\gamma_C$), so summing over $C \in \mathcal A_K$ makes $\hat F^\alpha$ weakly submodular with ratio $\gamma' = \min_{C \in \mathcal A_K} \gamma_C$. Let $S^*$ with $|S^*| = b$ satisfy $\min_{C\in\mathcal A_K} F_C(S^*) \ge \alpha$, i.e. $\hat F^\alpha(S^*) = |\mathcal A_K|\alpha$. At greedy step $i$ with current set $S_{i-1}$, weak submodularity on $S^* \cup S_{i-1}$ together with the optimality of $S^*$ gives $$\sum_{u \in S^*}[\hat F^\alpha(S_{i-1}\cup\{u\}) - \hat F^\alpha(S_{i-1})] \ge \gamma'(|\mathcal A_K|\alpha - \hat F^\alpha(S_{i-1})),$$ so the best single user obeys $\hat F^\alpha(S_{i-1}\cup\{u_i\}) - \hat F^\alpha(S_{i-1}) \ge \tfrac{\gamma'}{b}(|\mathcal A_K|\alpha - \hat F^\alpha(S_{i-1}))$. Writing the residual gap $R_i = |\mathcal A_K|\alpha - \hat F^\alpha(S_i)$, this is the recurrence $R_i \le R_{i-1}(1 - \gamma'/b)$, so from $R_0 = |\mathcal A_K|\alpha$ we get $R_N \le |\mathcal A_K|\alpha\, e^{-\gamma' N/b}$. Requiring $R_N \le \varepsilon$ yields $N > \tfrac{b}{\gamma'}\ln(|\mathcal A_K|\alpha/\varepsilon)$, hence the returned set satisfies $\min_{C \in \mathcal A_K} F_C(\hat S) \ge \alpha - \varepsilon$. Since one user is added per step and $\lceil x\rceil \le 1+x$, the seed-set size is bounded by $$|\hat{S}| \le  \left( 1 + \frac{1}{\gamma'} \log \left ( \frac {K \alpha} {\varepsilon} \right ) \right) b, \quad \text{where} \quad \gamma' = \min_{C \in \mathcal A_K} \frac {1} {\kappa(M \odot C)}.$$

\subsection{Proof of \cref{theorem:saturate-final}}

By \cref{lem:saturate-seeding}, the inner oracle at iteration $K$ returns $\hat S_K$ with $\min_{C \in \mathcal{A}_{K-1}} F(\hat{S}_K, L, C) \ge \alpha - \varepsilon$. The separation oracle then computes the global worst case $\hat C_K = \arg\min_{C \in \mathcal U} F(\hat S_K, L, C)$, and the termination condition $F(\hat S_K, L, \hat C_K) \ge \min_{C \in \mathcal{A}_{K-1}} F(\hat S_K, L, C) - \varepsilon_C$ yields $\min_{C \in \mathcal U} F(\hat S_K, L, C) = F(\hat S_K, L, \hat C_K) \ge (\alpha - \varepsilon) - \varepsilon_C = \alpha - (\varepsilon + \varepsilon_C)$.

\subsection{Proof of \cref{lemma:reg-seed}}

The regularized objective is $g(s+\delta_S, L, C) + \nu \|\delta_S\|^2 = (s + \delta_S)^T Z (s + \delta_S) + \nu \delta_S^T \delta_S$, whose gradient in $\delta_S$ vanishes at $Z_{SS}\delta_S + Z_{S:}s + \nu\delta_S = 0$, giving the unique minimizer $\delta_S^*(\nu) = -(Z_{SS}+\nu I)^{-1}Z_{S:} s.$
Substituting back, by the same algebra as \cref{lemma:submodularity} with $Z_{SS}$ replaced by $Z_{SS}+\nu I$, gives $F_\nu(S,L,C) = s^T P_S^\nu s$ with $P_S^\nu = Z_{:S}(Z_{SS}+\nu I)^{-1}Z_{S:}$, and weak submodularity follows identically with ratio $\gamma_\nu = \tfrac{\lambda_{\min}(Z)+\nu}{\lambda_{\max}(Z)+\nu} = \tfrac 1 {\kappa(Z+\nu I)}$. Finally, $\|\delta_S^*(\nu)\| \le 2/\nu$, since $\|(Z_{SS}+\nu I)^{-1}\| \le 1/\nu$ and $\|Z_{S:}s\| \le \|Z_{SS}\|\,\|s_S\| + \|Z_{S\bar S}\|\,\|s_{\bar S}\| \le 2$ due to the eigenvalue interlacing theorem.

\section{Helper Lemmas} \label{app:helper_lemmas}

\begin{lemma}
\label{lemma:rankone-marginal}
Let $Z\in\mathbb R^{n\times n}$ be symmetric and positive semidefinite.
Fix an index set $S\subseteq[n]$ for which $Z_{SS}$ is invertible, and
let $u\in[n]\setminus S$. Define
\begin{equation*}
P_S = Z_{:S} Z_{SS}^{-1} Z_{S:},\qquad
q_S = Z_{:u} - Z_{:S} Z_{SS}^{-1} Z_{Su},\qquad
\beta_S = Z_{uu} - Z_{uS} Z_{SS}^{-1} Z_{Su}.
\end{equation*}
Then $\beta_S>0$ and $P_{S\cup\{u\}} - P_S  =  \frac{q_S q_S^T}{\beta_S}.$

\end{lemma}

\xpar{Proof} Since $Z_{SS}\succ0$, the Schur complement of $Z_{uu}$ in the principal block of $Z$ on $S\cup\{u\}$ is $\beta_S = Z_{uu} - Z_{uS}Z_{SS}^{-1}Z_{Su}$, which is positive because the $2\times2$ block is PSD with $Z_{SS}\succ0$. The block-inverse formula gives
\begin{equation*}
\begin{pmatrix} Z_{SS} & Z_{Su} \\ Z_{uS} & Z_{uu}\end{pmatrix}^{-1} = \begin{pmatrix} Z_{SS}^{-1} + Z_{SS}^{-1}Z_{Su}\beta_S^{-1}Z_{uS}Z_{SS}^{-1} & -Z_{SS}^{-1}Z_{Su}\beta_S^{-1} \\ -\beta_S^{-1}Z_{uS}Z_{SS}^{-1} & \beta_S^{-1} \end{pmatrix}.
\end{equation*}
Expanding $P_{S\cup\{u\}} = Z_{:(S\cup\{u\})}\, Z_{S\cup\{u\}, S\cup\{u\}}^{-1}\, Z_{(S\cup\{u\}):}$ and collecting terms, the leading term is $P_S = Z_{:S}Z_{SS}^{-1}Z_{S:}$, while the remaining terms factor as $\beta_S^{-1} q_S q_S^T$ with $q_S = Z_{:u} - Z_{:S}Z_{SS}^{-1}Z_{Su}$. Hence $P_{S\cup\{u\}} = P_S + \beta_S^{-1} q_S q_S^T$, as claimed.



\section{Extensions and Generalizations}
\label{app:generalizations}
 
\subsection{Overlapping Groups}
\label{app:gen-overlap}
 
To accommodate overlap we replace the hard membership indicators of the base model by \emph{soft membership} vectors $\mu^{(k)}\in[0,1]^n$, where $\mu^{(k)}_i$ is the degree to which user $i$ belongs to group $k$, with the fuzzy-partition normalization $\sum_k \mu^{(k)} = \one$. The consensus attributable to group $k$ is $z_k  =  (I+L)^{-1}\left (\mu^{(k)}\odot s\right )$. For two (possibly overlapping) groups $k,\ell$, write
$d^{(k\ell)}=\mu^{(k)}-\mu^{(\ell)}$ and $y^{(k\ell)}=d^{(k\ell)}\odot s$. The
conditional disparity becomes

\begin{equation}
  f(s,L,k,\ell)
   =  \big\|z_k-z_\ell\big\|^2
   =  \left (y^{(k\ell)}\right )^T (I+L)^{-2}  y^{(k\ell)}
   =  \left (y^{(k\ell)}\right )^T M  y^{(k\ell)},
  \label{eq:overlap-cond}
\end{equation}

which is structurally identical to the hard-partition expression $f(s,L,A)=y^T M y$ of \cref{sec:disparity}, the only change being that the assignment vector $r$ is replaced by the real-valued vector $d^{(k\ell)}=\mu^{(k)}-\mu^{(\ell)}\in[-1,1]^n$.
 
\begin{proposition}
\label{prop:overlap-axioms}
For each group $k$, the soft-attribution map $\Phi(s,k)=(I+L)^{-1}\left (\mu^{(k)}
\odot s\right )$ is linear in $s$, and the family $\{\Phi(\cdot,k)\}_k$ satisfies
additivity $\sum_k \Phi(s,k)=(I+L)^{-1}s$ whenever $\sum_k\mu^{(k)}=\one$, and affine invariance with respect to the opinions of any complementary group.
\end{proposition}
 
The argument is unchanged from \cref{app:definition}: that proof uses only the linearity of $s\mapsto\Phi(s,k)$ and additivity, never that the weights are $\{0,1\}$-valued, and affine invariance again forces a zero baseline for the unobserved group. 
 
Moreover, the platform's classifier now outputs \emph{soft scores} rather than hard labels. Centering membership as $\tilde r^{(k)}_i = 2\mu^{(k)}_i-1\in[-1,1]$, the relevant group structure is $C^{(k)}_{ij}= \ev {} {\tilde
r^{(k)}_i \tilde r^{(k)}_j}$ with $\diag (C^{(k)}) \preceq I$. The average disparity retains its quadratic form $g(s,L,C^{(k)})=s^T \left (M\odot C^{(k)}\right )s$, and the uncertainty set
$\mathcal U(\rho)$ is defined entrywise exactly as in \cref{sec:interventions} so the robust recommendation-reweighing and seeding problems can be extended similarly. 
 
\subsection{Multiple Groups}
\label{app:gen-multi}
 
Let the population be (softly or hardly) partitioned into $K\ge 2$ groups with attributed consensuses $z_1,\dots,z_K$, where $z_k=(I+L)^{-1}(\mu^{(k)}\odot s)$ and $\sum_k z_k = z$. The binary disparity compares the two attributions directly; with $K$ groups we have $\binom{K}{2}$ pairwise comparisons, and there are two natural ways to aggregate them.
 
On the one hand, we may fix nonnegative weights $w_{k\ell}\ge 0$ (e.g.\ $w_{k\ell}=1$ for the uniform
choice, or $w_{k\ell}\propto |A_k| |A_\ell|$ to weight by group size) and define $$\mathcal D(s,L,\{\mu^{(k)}\})
   =  \sum_{k<\ell} w_{k\ell}  f(s,L,k,\ell)
   =  \sum_{k<\ell} w_{k\ell} \left (y^{(k\ell)}\right )^T M  y^{(k\ell)}.
$$
Each summand is a quadratic in $s$ through the shared kernel $M$, so the aggregate is again a quadratic form $s^T Z_{\mathrm{multi}} s$ with $Z_{\mathrm{multi}}$ a PSD matrix built from $M$ and the weighted pairwise contrasts; the structural disparity is $\lambda_{\max}(Z_{\mathrm{multi}})$, and the intervention algorithms of \cref{sec:interventions} apply with $Z$ replaced by $Z_{\mathrm{multi}}$. Alternatively, when the platform's concern is the single most over-represented group rather than an average, the optimization problem becomes optimizing $g_{\mathrm{multi}}(s,L)
   =  \max_{k\neq \ell}  g\left (s,L,C^{(k\ell)}\right )$, with $C^{(k\ell)}_{ij} = \mathbb E \big[\tilde r^{(k)}_i  \tilde r^{(\ell)}_j\big]$. The platform faces a per-pair uncertainty set $\mathcal U^{(k\ell)}(\rho)$ and solves $\min_{L\in\mathcal L} 
  \max_{k\neq\ell}
  \max_{C\in\mathcal U^{(k\ell)}(\rho)} 
  \lambda_{\max}\left (M\odot C\right )$, i.e., a single outer minimization against a worst case taken jointly over the group pair and the correlation structure. The inner maximum over a finite set of
pairs preserves convexity of the surrogate objective, so the active set algorithms extend by enlarging the active set to range over $(k,\ell)$ pairs in addition to candidate correlation matrices.
 
 
\subsection{General Linear Consensus Models}
\label{app:gen-linear}
 
The FJ map $z=(I+L)^{-1}s$ is just one instance; the metric is really a property of the consensus operator and extends to any \emph{linear} model $z=Ts$. For a partition with contrast $r$ and $y = r\odot s$, the conditional, average, and structural disparities become $f_T(s,T,A) = \|T(\one_A\odot s)-T(\one_{\bar A}\odot s)\|^2 = y^T M y$, $g_T(s,T,C) = s^T(M\odot C)s$, and $h_T(T,C) = \lambda_{\max}(M\odot C)$, with $M = T^T T$. The main-text results carry over whenever $T$ is \emph{(T1)} linear in $s$, \emph{(T2)} satisfies $M = T^T T \succeq 0$, and \emph{(T3)} row-stochastic ($T\one=\one$).

\xpar{Examples} Several standard models satisfy (T1)--(T3). For \emph{agents with varying susceptibility to persuasion} $\gamma > 0$ \citep{Abebe2018}, $z = Ts$ with $T = \gamma\big((1-\gamma)L + \gamma I\big)^{-1}$, whose eigenvalues are $\gamma/((1-\gamma)\lambda_i + \gamma)$, so all our algorithms transfer directly. For the \emph{DeGroot model} \citep{DeGrootModel}, $z = (q^Ts)\one$ gives $T = \one q^T$ and $M = n qq^T$, where $q$ is the stationary vector of the chain $D^{-1}A$, i.e.\ $q_i = d_i/2m$; the link-optimization variable is then the degree vector over $\mathcal D = \{d : d_i \ge 0,\ \sum_i d_i = 2m\}$.

\xpar{Leverage scores} If $T$ is differentiable in the edge weights, the robust leverage scores generalize to $\beta_e^{\mathrm{rob}, T} = v^T(\tfrac{\partial M}{\partial w_e}\odot C^*)v$, where $C^*$ is the worst-case group structure and $v$ the leading eigenvector of $M\odot C^*$. For the DeGroot model in particular, $\beta_e^{\mathrm{rob}, \dg} = \tfrac{n}{m}(\psi_i + \psi_j - 2q^T\psi)$ for $e=(i,j)$, with $\psi = s\odot(C(s\odot q))$.

\end{APPENDIX}

\end{document}